\documentclass[fleqn,usenatbib]{mnras}

\usepackage{newtxtext,newtxmath}

\usepackage[T1]{fontenc}

\DeclareRobustCommand{\VAN}[3]{#2}
\let\VANthebibliography\thebibliography
\def\thebibliography{\DeclareRobustCommand{\VAN}[3]{##3}\VANthebibliography}

\usepackage{graphicx}	
\usepackage{amsmath}	
\usepackage{booktabs}
\newcommand{\RN}[1]{%
  \textup{\uppercase\expandafter{\romannumeral#1}}%
}

\title[Simulating HVCs in the Observational Plane]{Simulating High-Velocity Clouds in the Observational Plane: An Initial Study with the Smith Cloud}

\author[L. E. Porter]{
Lori E. Porter,$^{1}$\thanks{E-mail: lep2176@columbia.edu}
Matthew Abruzzo,$^{2}$
Greg L. Bryan,$^{1}$
Mary Putman,$^{1}$
Yong Zheng,$^{3}$
and Drummond Fielding$^{4}$
\\
$^{1}$Department of Astronomy, Columbia University, 538 West 120 Street, New York, NY 10027, USA\\
$^{2}$Department of Physics and Astronomy, University of Pittsburgh, 3941 O'Hara St, Pittsburgh, PA 15260, USA\\
$^{3}$Department of Physics, Applied Physics and Astronomy, Rensselaer Polytechnic Institute, Troy, NY 12180, USA \\
$^{4}$Department of Astronomy, Cornell University, Ithaca, NY 14853, USA
}

\date{Accepted XXX. Received YYY; in original form ZZZ}

\pubyear{\the\year}

\begin{document}
\label{firstpage}
\pagerange{\pageref{firstpage}--\pageref{lastpage}}
\maketitle

\begin{abstract}
High-velocity clouds (HVCs) may fuel future star formation in the Milky Way, but they must first survive their passage through the hot halo. While recent work has improved our understanding of the survival criterion for cloud-wind interactions, few observational comparisons exist that test this criterion. We therefore present an initial comparison of simulations with the Smith Cloud (SC; $d=$ 12.4 kpc, $l, b = 40^{\circ}, -13^{\circ}$) as mapped with the GALFA-H\MakeUppercase{\romannumeral1} survey. We use the Smith Cloud's observed properties to motivate simulations of comparable clouds in wind tunnel simulations with Enzo-E, an MHD code. For both observations and simulations, we generate moment maps, characterize turbulence through a projected first-order velocity structure function (VSF), and do the same for H\MakeUppercase{\romannumeral1} column density with a normalized autocovariance function. We explore how initial cloud conditions (such as radius, metallicity, thermal pressure, viewing angle, and distance) affect these statistics, demonstrating that the small-scale VSF is sensitive to cloud turbulence while large scales depend on cloud bulk velocity and viewing angle.
We find that some simulations reproduce key observational features (particularly the correlation between column density and velocity dispersion) but none match all observational probes at the same time (the large scales of the column density autocovariance is particularly challenging). We find that the simulated cloud (cloud C) showing growth via a turbulent radiative mixing layer (TRML) is the best match, implying the importance of TRML-mediated cooling for Milky Way HVCs.
We conclude by suggesting improvements for simulations to better match observed HVCs.  
\end{abstract}




\section{Introduction}


The accretion of gas onto galaxies, including our own Milky Way (MW), is essential to understand. With our own Galaxy having a current star formation rate (SFR) of a few solar masses per year, a lack of gas accretion could mean the Milky Way will deplete its gas reservoir in the next billion years if it maintains the current SFR \citep{Chiappini2001, Chiappini2003, Fuchs2009, Robitaille2010, Chomiuk2011, Kennicutt2012, Putman2012, Licquia2015, Elia2022}. Therefore, gas accretion is required to provide a continuous source of fuel for the Galaxy's future star formation \citep{Erb2008, Hopkins2008, Kennicutt2012, Putman2012, Fox2019}. 

As the coldest and densest component of halo gas, high-velocity clouds (HVCs) are thought to be a potential source of future star formation \citep{Oort1969, Larson1980, Wakker1997, Putman2003, Joung2012, Putman2012, Fox2014, Henley2017, Fox2019, Lehner2022}. Hundreds of HVCs have been observed across the entire sky, with many belonging to larger complexes or filamentary structures, and are typically defined by their higher local-standard-of-rest velocities ($|v_{\rm LSR}|>70-90$ km/s), or the deviation velocity from Galactic rotation ($|v_{\rm dev}|>75$ km/s). Due to their ubiquity across the sky, HVCs have a wealth of observational data available at various resolutions (see \citealt{Wakker1997, Putman2012} and references therein for comprehensive reviews).

All potential new Galactic fuel sources, including HVCs, must pass through a galaxy's halo without becoming homogenized with the hot halo, but surviving this passage depends on a wide variety of factors. Because observations of H\MakeUppercase{\romannumeral1} structure in these clouds is thought to be evidence that they are moving through the Galaxy's diffuse halo medium and being disrupted \citep{Bruns2000, Sembach2003, Putman2003, Maller2004, Stanimirovic2006, Peek2007, Putman2012}, the question of whether HVC survive passage through the Galactic halo can be modeled as a cool cloud embedded within a hot medium setup, or ``wind tunnel'' framework \citep{Heitsch2009, Heitsch2016, Heitsch2022, Henley2017, Bustard2022, Tan2023}. 

The generic scenario where a pressure-confined, cool cloud ($10^{3-4}$ K) moves with respect to a coherent hotter ($>10^6$ K), volume-filling background flow (or ``wind'') has been studied for decades \citep{McKee1977, Balbus1982}. It has received significant attention in the context of explaining the origin of multiphase galactic outflows, which are commonly composed of comoving cooler and hotter gas phases \citep[e.g][]{Pillepich2018, Dave2019}. 
The conventional model holds that supernovae launch hot winds that subsequently accelerate and entrain cool clouds (historically, via ram pressure acceleration) encountered in the ISM as the flow propagates out of the galaxy.
However, simulating cloud survival and acceleration has proven remarkably difficult \citep[e.g.][]{Cooper2009, Scannapieco2015, Schneider2017, Sparre2019} because the initial velocity differential drives hydrodynamical instabilities and turbulent mixing that destroys the cloud. 
A cloud of initial radius $R_{\rm cl}$ is homogenized with a background flow (of speed $v_{\rm w}$ and initial density contrast $\chi=\rho_{\rm cl}/\rho_{\rm w}$) over a few cloud-crushing times $t_{\rm cc} = \chi^{1/2}\frac{R_{\rm cl}}{v_{\rm w}}$ \citep{Klein1994}, which is shorter than the ram pressure acceleration timescale, $\chi \frac{R_{\rm cl}}{v_{\rm w}}$  \citep{Zhang2017a, Abruzzo2022}. 

Only recently have simulations been able to model cloud survival in the regime of rapid radiative cooling \citep{Marinacci2010, Armillotta2016, Gronke2018}. 
\citet{Gronke2018} showed that when intermediate temperature gas produced by mixing cools sufficiently quickly, mixing acts as a mechanism for transferring mass and momentum to the cloud from the hot phase rather than destroying the cloud. 
In this regime, the cloud not only survives, but grows. 
We refer to this process as TRML (turbulent radiative mixing layer) entrainment, a phenomenon whose underlying physics and significance have motivated extensive simulations \citep[e.g.][]{Ji2019,Fielding2020, Gronke2020a, Tan2021, Abruzzo2022, Bustard2022, Chen2023b, Chen2024,Hidalgo-Pineda2024, Richie2024}.

The criteria for cloud survival has been a topic of debate that largely arises from diverging $\chi\gtrsim300$ simulation results \citep[e.g][]{Gronke2018,Li2020b,Sparre2020,Kanjilal2021,Farber2022}.
\citet{Abruzzo2023} resolved this debate by finding that diverging results originated from different choices of cooling functions, whose shapes are known to impact cloud survival \citep{Abruzzo2022}, and proposed a unified survival criterion.
The criterion derives from the observation that clouds are destroyed unless they grow, and for a cloud to grow, mixed material originating from the hot phase must be absorbed into the cloud.
Thus, mixed hot phase material must cool to the cloud temperature before advecting past the end of the cloud (at which point mixing with the background heats the material back up). This can be restated as $t_{\rm cool,minmix} < \alpha t_{\rm sh}$, where $t_{\rm cool,minmix}$ estimates the characteristic cooling timescale of mixed gas \citep[adapted from][]{Farber2022} and $\alpha t_{\rm sh}$ estimates the advection timescale. $t_{\rm sh}=R_{\rm cl}/v_{\rm w}$ specifies the time for pristine hot phase material to advect the initial length of the cloud, and $\alpha$ is an empirical constant (${\sim}7$) that accounts for the cloud elongation, entrainment, and reduction in speed of hot phase material from mixing.
This criterion can be recast as a minimum cloud survival radius, $R_{\rm crit} = v_{\rm w} t_{\rm cool,minmix} \alpha^{-1}$.


Despite new progress in understanding the cloud survival criterion and the relevant physical parameters, many of the quantities in theoretical studies are difficult to determine observationally (Mach number, thermal pressure, etc.). As a result, observational tracers of cloud survival or destruction need to be identified in order to test these criterion on observed clouds. Milky Way HVCs then become ideal probes as their proximity allows detailed spatial and kinematic structure to be observed \citep{Wakker2002, BenBekhti2009, Hsu2011}.

This motivation has driven various simulations and models of HVCs \citep{Heitsch2009, Heitsch2016, Henley2017, Bustard2022, Heitsch2022, Tan2023}. For instance, \citet{Bustard2022} investigate survival criterion and TRML in the Magellanic System, predicting that the Leading Arm and part of the trailing Magellanic Stream should survive and even gain mass. \citet{Tan2023} emphasize the importance of gravity, suggesting that survival of HVCs is dependent on the radius of the cloud and dropping height. They predict clouds of height $\leq 10$ kpc from the MW to survive their infall, while clouds with larger heights require larger radii to survive \citep[see also][]{Heitsch2009, Lehner2022}. While they do not specifically investigate HVCs, \citet{Gronke2022} and \citet{Abruzzo2024} both investigate how the velocity structure function (VSF) could be used to characterize turbulence over various length scales $\ell$, and further relate the behavior of turbulence to different stages of the cloud survival process.
However, these previous studies have been limited in regards to what information can be \emph{directly} used and compared from \emph{both} observations and simulations. In short, there are currently no systematic and comprehensive comparisons of HVCs and simulations in the \emph{observational plane}. 

One of the most well-studied Milky Way HVCs is the Smith Cloud (SC). First discovered by \citet{Smith1963}, the SC is a relatively large HVC in a high-pressure environment, with the head located at $(l, b) = (40^\circ, -13^\circ)$, and total length of about 20$^\circ$, inferred to be moving across the sky with an angle of $45^\circ\pm10^\circ$ \citep{Lockman2008}. Distance estimates for the SC are about 12.4 kpc from the sun, and within 3 kpc of the Galactic plane, falling towards it at a rate of about $v_z\sim 70 \rm \; km \; s^{-1}$ \citep{Putman2003, Lockman2008, Wakker2008, Fox2016}. It has a metallicity of half solar \citep{Fox2016} and an H\MakeUppercase{\romannumeral1} mass of $10^6 M_{\odot}$, with total mass of about $2\times10^6 M_{\odot}$ \citep{Lockman2008, Hill2009}. The SC also has an unusually high peak H\MakeUppercase{\romannumeral1} column density of $N_{\rm H\MakeUppercase{\romannumeral1}}\approx 3\times10^{20} \rm cm^{-2}$  compared to most HVCs (peak $N_{\rm H\MakeUppercase{\romannumeral1}}\sim 10^{19} \rm cm^{-2};$ \citealt{Putman2012}). Due to the Smith Cloud's position in the lower halo, it is expected to reflect a surviving cloud. 

With the well-constrained properties of the Smith Cloud from new high-resolution GALFA-HI observations and recent advances in simulations of cloud-wind interactions, we present an initial test comparing observed HVCs to simulated (mock observed) clouds in a systematic and consistent manner.  In addition, we vary the initial conditions of the simulated clouds, including metallicity, radius, pressure, cloud orientation, and distance in order to determine the effect of each parameter on our measurements.
This paper is organized as follows: In Section~\ref{observationdetails} we outline the GALFA-HI observations, Section~\ref{simulationinfo} details the cloud-wind simulations used, and Section~\ref{results} provides an overview of results on observations and mock observations (simulations), including moment maps (\ref{momentmaps}), the projected first-order velocity structure function (\ref{vsf}), the normalized autocovariance function of column density (\ref{acf}), and a discussion of how cloud properties may impact these statistics (\ref{cloudorientation_clouddistance}). We discuss our results and caveats in comparison with previous literature in Section~\ref{discussion}, and provide our conclusions and plans for future work in Section~\ref{conclusions}.

\section{Observations}\label{observationdetails}

We conduct our observational analysis using data from the Galactic Arecibo L-Band Feed Array HI (GALFA-HI) survey, detailed by \citet{Peek2018}. GALFA-HI boasts a high sensitivity and resolution of 4', equivalent to a physical size of approximately 14.4 pc at the distance of the Smith Cloud (12.4 kpc), resulting in the most resolved observations of the Smith Cloud (actual size about 1 kpc x 3 kpc) to date. 

The GALFA-HI cube of the Smith Cloud has a 0.739 $\rm km \; s^{-1}$ spacing between velocity channels across a total range of $-68 < v_{\rm LSR} < 188 \rm \; km \; s^{-1}$, as used in \cite{Holmhansen2025}.  We reduced the velocity range to $75 < v_{\rm LSR} < 130 \rm \; km \; s^{-1}$, capturing the majority of Smith Cloud emission while avoiding Galactic contamination.  The observations cover the area of the Smith Cloud from $295^{\circ} < \alpha < 312.5^{\circ}$ and $-0.725^{\circ} < \delta < 3.425^{\circ}$.  Due to the limit of the GALFA-HI declination range, we do not capture the entirety of the Smith Cloud, as the main component extends down to $\delta\sim-2^{\circ}$ \citep[e.g.,][]{Lockman2008}, which we address in Section~\ref{caveats}. 

\begin{figure}
    \centering
    \includegraphics[width=\columnwidth]{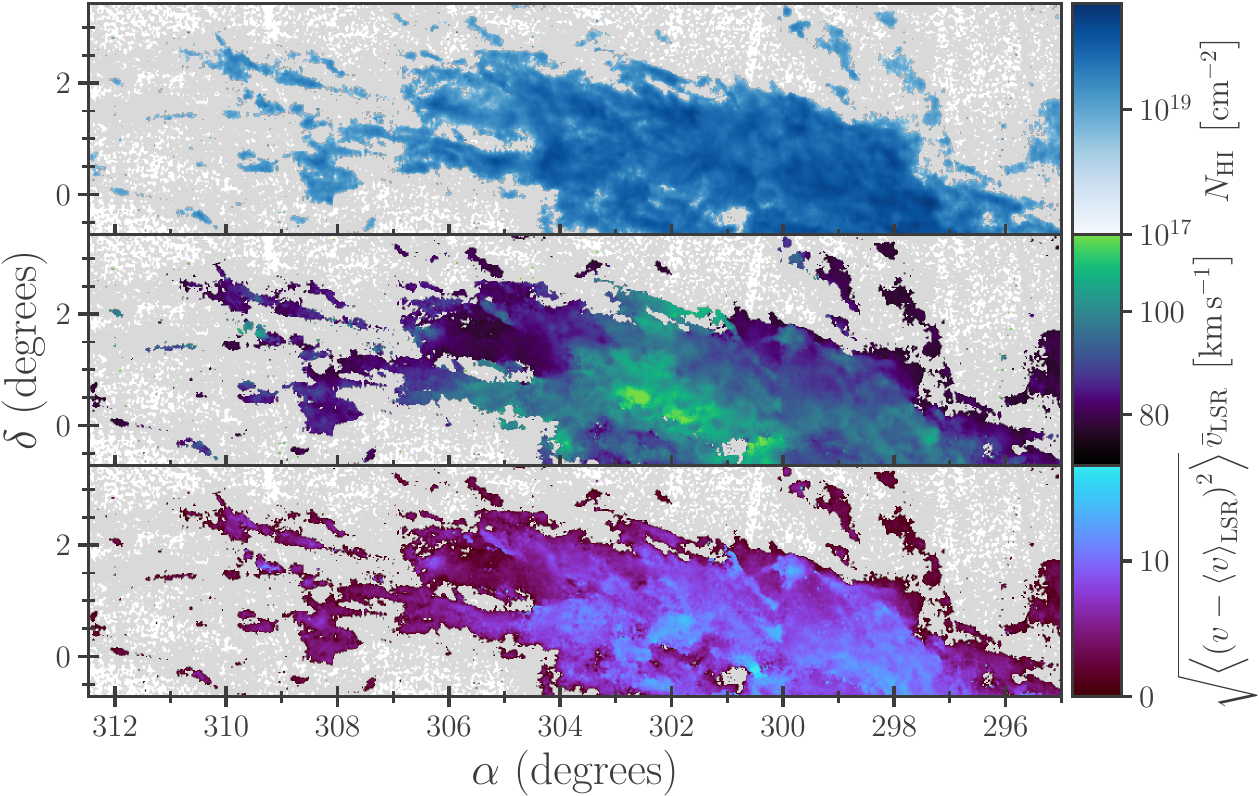}
    \caption{GALFA-H\MakeUppercase{\romannumeral1} moment maps of the Smith Cloud, smoothed over 4 km/s in spectral resolution, then integrated across 75-130 $\rm km \; s^{-1}$ and with a 2.5$\sigma$ clipping data reduction. Grey regions represent parts of the observations that are removed by applying the 2.5$\sigma$ clipping to brightness temperature when calculating spectral moments.  Top row shows the zeroth moment (neutral hydrogen column density; $N_{\rm H\MakeUppercase{\romannumeral1}}$), second row is the intensity-weighted average $\rm v_{\rm LSR}$, and third row is the intensity-weighted velocity dispersion.}
    \label{fig:observations_smoothed_2sigma}
\end{figure}

To improve the signal-to-noise ratio, we smooth the observations over 4 $\rm km \; s^{-1}$ in the spectral dimension. We also employ sigma-clipping with a $2.5\sigma$ limit on brightness temperature when calculating the spectral moments.  The observational data after smoothing and sigma clipping is visible in Figure~\ref{fig:observations_smoothed_2sigma}, resulting in a minimum column density $N_{\rm HI,min}\approx6.61\times10^{17} \, {\rm cm^{-2}}$ and median $N_{\rm HI,med}\approx 4.66\times10^{19} \, {\rm cm^{-2}}$.

\section{Simulations}\label{simulationinfo}

\begin{table*}\caption{Initial conditions (ICs) for each simulation run, including the thermal pressure ($p/k_B$), radius of the cloud ($R_{\rm cl}$; chosen such that the mass $M_{\rm cl}$ is $10^6 M_{\odot}$), metallicity of the cloud ($Z_{\rm cl}$), temperature and velocity of the wind ($T_{\rm w}$, $v_{\rm w}$), Mach number ($\mathcal{M}$), and density contrast between the cloud and wind ($\chi = \rho_{\rm cl}/\rho_{\rm w}$). We also include whether the simulated cloud eventually survives, and any additional notes on the particular simulation. Each simulation here has Mach number $\mathcal{M}=1.5$, resolution of $R_{\rm cl}/\Delta x = 16$, and heating and cooling are shut off for $T_{\rm cl} > 0.6 T_{\rm w}$.}
\label{table:sim_parameters}
    \begin{tabular}{lcccccccccc} 
\hline
 & $p/k_B \; {\rm \left[K \, cm^{-3}\right]}$ & $R_{\rm cl} \; {\left[\rm pc\right]}$  &  $Z_{\rm cl} \; {\rm \left[Z_{\odot}\right]}$ & $T_{\rm w} \; {\rm \left[K\right]}$ & $v_{\rm w} \; {\rm \left[km \, s^{-1}\right]}$ & $\chi$ & Survival? & Notes \\ 
\hline
Simulation A & $10^3$ & 445 & 1 & $1.71\times10^6$ & $292$ & 300 & Yes & Heating/cooling shut off below $T_{cl}$ \\

Simulation B & $10^3$ & 445 & 0.5 & $1.71\times10^6$ & $292$ & 300 & Yes & Heating/cooling shut off below $T_{cl}$ \\

Simulation C & $5\times10^3$ & 169 & 0.5 & $7.49\times10^5$ & $193$ & 300 & Yes & Heating/cooling normal below $T_{cl}$ \\

Simulation D$^{*}$ & $5\times10^3$ & 169 & 0.5 & $7.49\times10^5$  & $111$ & 100$^{**}$ & No & No cooling \\
\hline
\hline
\multicolumn{9}{p{\textwidth}}{$^*$Properties have been re-scaled from an existing simulation in the library from \citet{Abruzzo2024}. }\\
\multicolumn{9}{p{\textwidth}}{$^{**}$ An additional simulation with the same ICs as simulation D, but with $\chi=1000$, was run to ensure the $\chi$ difference compared to $\chi=300$ of A-C did not result in significant differences. Because the $\chi=1000$ adiabatic simulation produces comparable observational results to simulation D ($\chi=100$), we do not show the results here and conclude that this should not affect comparison: a similar $\chi=300$ adiabatic simulation would have minimal differences with simulation D. }\\
    \end{tabular}
\end{table*}

Following \citet{Abruzzo2024}, clouds are simulated in a `wind tunnel' setup by running a suite of 3D uniform grid hydrodynamical simulations using Enzo-E\footnote{\url{http://enzo-e.readthedocs.io}}, a rewrite of Enzo \citep{Bryan2014}, which is built on the adaptive mesh refinement framework CELLO \citep{BordnerNorman2012, BordnerNorman2018}.

We conduct four total simulations, named alphabetically as simulations A, B, C, and D, respectively. All simulated clouds begin as a slightly perturbed spherical cloud with no initial velocity, embedded within a hot, uniform, laminar wind in the $+\hat{x}$ direction. In an attempt to replicate observations of the Smith Cloud, simulation initial conditions are chosen based on observed properties of the SC, described below, though we vary a few parameters between simulations in order to determine how they affect results. Table~\ref{table:sim_parameters} lists the chosen parameter values for each simulation. All simulations here are run with Mach number $\mathcal{M}=1.5$, resolution of $R_{\rm cl}/\Delta x = 16$, and density contrast $\chi=300$, with the exception of simulation D, which has $\chi=100$.

We require that simulated cloud properties obey a simplistic model. The initial cloud temperature is determined by $T_{\rm cl} = f(p, Z_{\rm cl})$, where $f(p, Z_{\rm cl})$ provides the temperature for thermal equilibrium depending on the thermal pressure $p$ and choice of cooling curve, which is determined by cloud metallicity $Z_{\rm cl}$. Then, from the ideal gas law and  $\rho_{\rm cl}=n_{\rm cl}m_{\rm H}\mu(p,T_{\rm cl},Z_{\rm cl})$, where $\mu$ is the mean molecular weight and $m_{\rm H}$ is the mass of hydrogen, the density of the cloud is

\begin{equation}\label{clouddensity}
    \rho_{\rm cl} = \frac{p\, m_{\rm H} \mu(p, T_{\rm cl}, Z_{\rm cl})}{k_{\rm B}T_{\rm cl}}.
\end{equation}

Then, because the cloud is initialized as a sphere, the cloud radius is determined by $R_{\rm cl} = \left(\frac{3M_{\rm cl}}{4\pi \rho_{\rm cl}}\right)^{1/3}$.

Properties of the simulated clouds are determined by observations, such as $M_{\rm cl}\approx 10^6 M_{\odot}$ \citep{Lockman2008} and $Z_{\rm cl}=Z_{\odot}/2$ \citep{Fox2016}. As a result, the \textit{only} free parameter we are left with is thermal pressure. Our initial choice of $p/k_B = 10^3 \, {\rm K \, cm^{-3}}$ is motivated by this being a standard and reasonable assumption in prior cloud-wind simulations.

Finding that simulations A and B produced lower column densities ($N_{\rm cl}$) than the Smith Cloud prompted us to run simulation C. The relation $N_{\rm cl} \propto n_{\rm cl}^{2/3}$ (from $R_{\rm cl} \propto n_{\rm cl}^{-1/3}$ and $N_{\rm cl} \sim R_{\rm cl} n_{\rm cl}$) motivated our choice of conditions. We initialized simulation C with $p/k_B=5\times10^3 \; \rm K \; cm^{-3}$, $T_{\rm cl}=4430$ K, and $R_{cl}= 169$ pc to maintain mass and metallicity values similar to observations, while producing column densities that are higher by a factor of $\approx 4$.

Radiative cooling is modeled with the GRACKLE\footnote{\url{http://grackle.readthedocs.io}} library \citep{Smith2017}, assuming metallicity estimates of $Z_{\odot}$ for simulation A and $Z_{\odot}/2$ otherwise, and no self-shielding. As in \citet{Abruzzo2024}, this consists of the tabulated heating and cooling rates for optically thin gas in ionization equilibrium with the $z=0$ \citet{Haardt2012} UV background. Radiative cooling in simulations A and B is shut off below the initial cloud temperature of $T_{\rm cl}=7940$ K, while in simulation C, heating and cooling work normally below the initial cloud temperature, $T_{\rm cl}=4430$ K. In all simulations, heating and cooling is shut off for $T_{\rm cl} > 0.6 T_{\rm w}$ so that the background medium does not cool. 

To determine how radiative cooling --- or lack thereof --- affects the results, we re-scale the cloud radius, specific internal energy (and by extension, temperature), and thermal pressure of a simulation with no cooling from the suite by \citet{Abruzzo2024} to have the same properties of simulation C, resulting in simulation D. As noted in \citet{Abruzzo2022}, we are free to do this due to the self-similarity of the problem in the absence of cooling. We note that these simulations do not include self-shielding, gravity, or magnetic fields, which we defer to subsequent work and discuss the impacts of in Section~\ref{caveats} (see also caveats of \citealt{Abruzzo2024}).

To project clouds to the observational plane in the form of position-position-velocity cubes, we pick a location to place the cloud relative to the observer, guided by distance estimates for the Smith Cloud from observations \citep{Putman2003, Lockman2008, Wakker2008, Fox2016}. In lieu of a flat-sky approximation and in line with GALFA-HI observations, we construct an array of vectors sampled from a grid of right ascension and declination values, chosen to cover the same region as the GALFA-HI observations, and carry out radiative transfer calculations along each ray, accounting for the local HI density as well as Doppler line broadening from cell temperatures and Doppler shifts due to the component of the velocity along the ray. We estimated the HI density by combining the electron density derived from Grackle's tabulated mean-molecular weight calculations with the assumptions that the gas has a neutral charge and H is as ionized as possible. The resulting image from the plate carr\'ee (or equirectangular) projection is then convolved to have a beam response function to match the GALFA-HI observations. 

Once mock observational cubes are produced, we then add Gaussian noise to the simulations that roughly approximates the noise in the Smith Cloud observations over emission-free velocity channels, and smooth the simulations in the same way as observations, over 4 $\rm km \; s^{-1}$ in spectral resolution. Finally, we implement the same 2.5$\sigma$ clipping level as used for the observations.

\begin{figure*}
	\includegraphics[width=\textwidth]{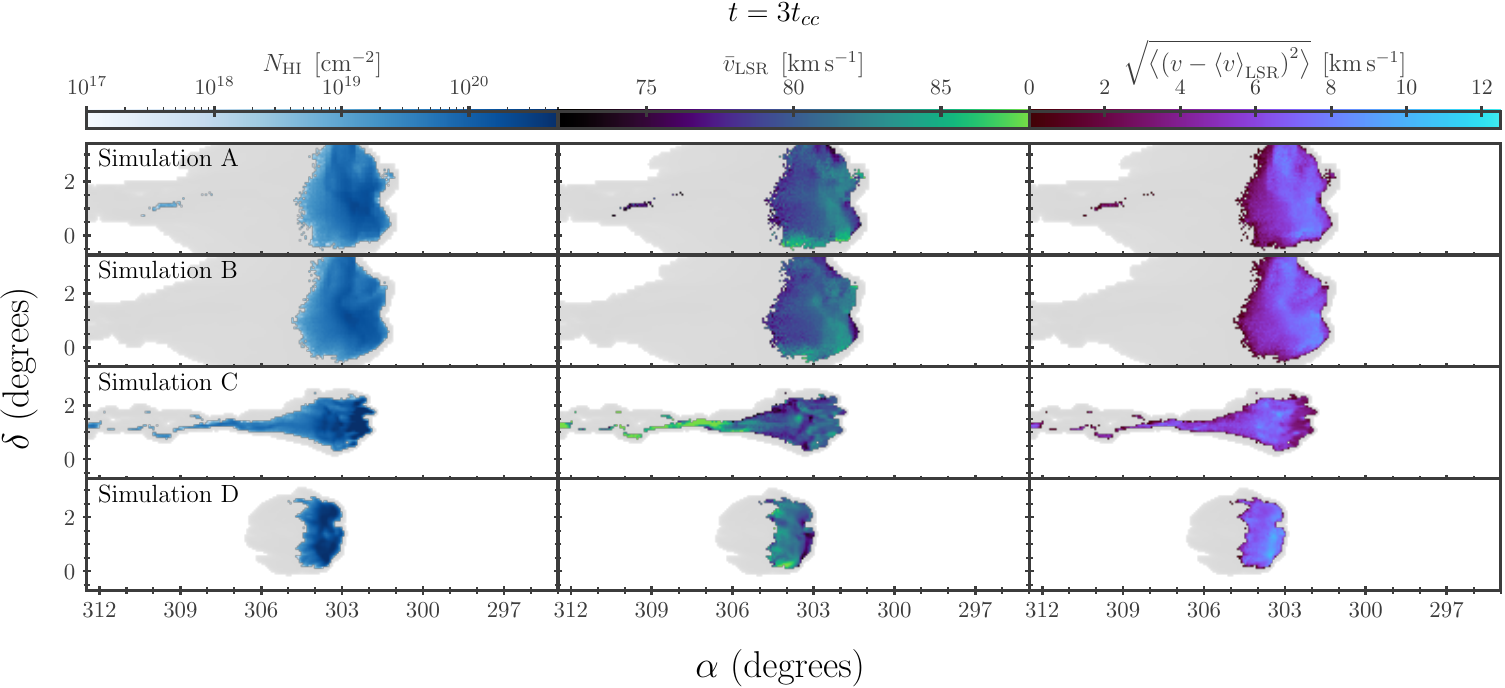}
    \caption{Moment maps of nearly all simulated clouds, with rows organized by simulation and columns by column density (column 1; zeroth moment), intensity-weighted mean velocity (column 2; first moment), and intensity-weighted velocity dispersion (column 3; second moment). All quantities are integrated across the entire simulations' velocity range at $t=3t_{\rm cc}$. Grey regions represent the entire simulated cloud (without the addition of noise), including data that is removed when applying the 2.5$\sigma$ clipping. Because both simulations C and D are initialized at the same radius, it becomes apparent that simulation D has already begun to become homogenized within the wind, therefore undergoing destruction. We note that these images do not cover the entirety of the simulation volume, and the simulation domain is chosen to minimize boundary effects.}
    \label{fig:sims_tcc03_2sigma}
\end{figure*}

\begin{figure*}
	\includegraphics[width=\textwidth]{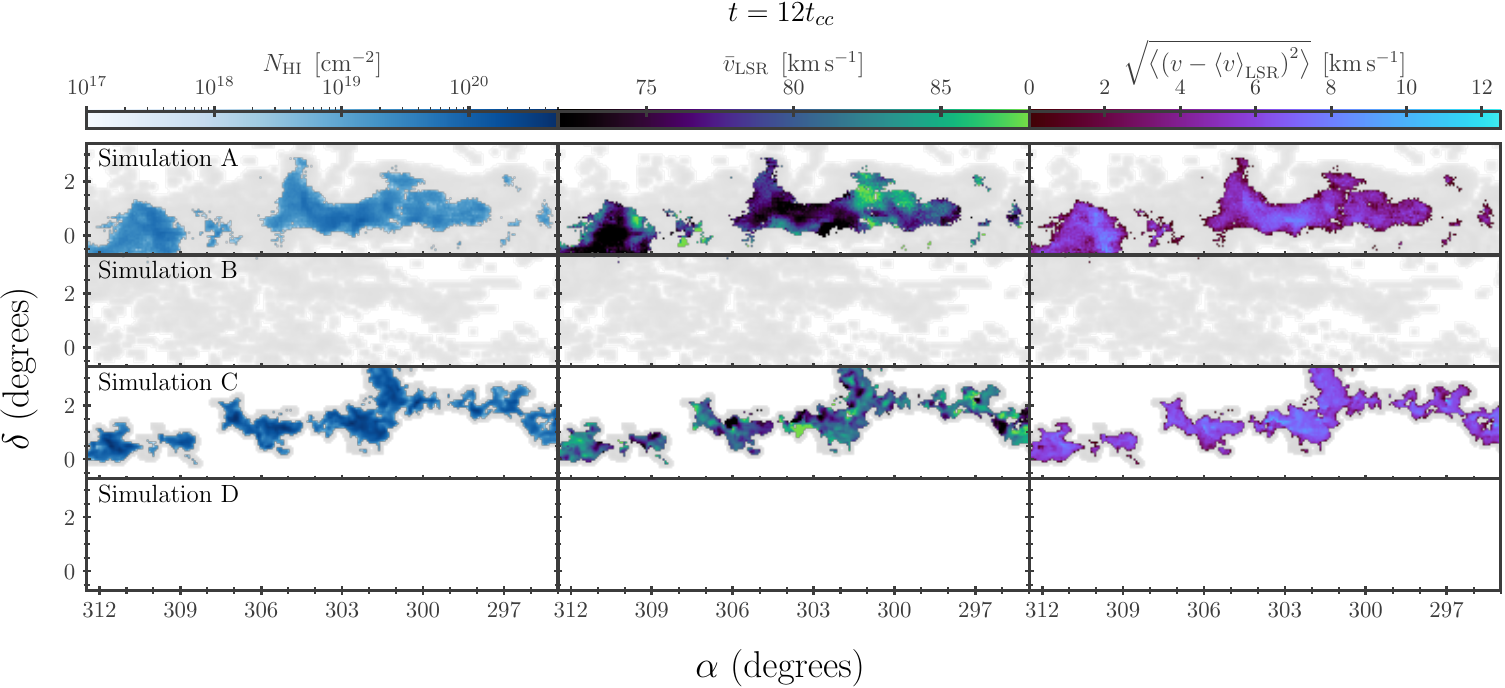}
    \caption{Same as Figure~\ref{fig:sims_tcc03_2sigma}, but at $t = 12t_{\rm cc}$. Simulation D does not survive the cloud-wind interaction and is homogenized with the wind (destroyed) by $6t_{\rm cc}$. Simulations A and C have been entrained in the wind and are growing, while simulation B has no visible data after the $2.5\sigma$ clipping. Simulation D is homogenized with the background wind, representing our definition of a destroyed cloud. As in Figure~\ref{fig:sims_tcc03_2sigma}, this does not represent the entire simulation volume, but the domain is selected to minimize boundary effects.}
    \label{fig:sims_tcc12_2sigma}
\end{figure*}

\section{Results}\label{results}

To make comparisons with the Smith Cloud, we contrast properties that are derived from the SC observations with mock observations of our simulations in identical fashion.
We focus on moment maps (and by extension, spatial morphology), velocity structure functions, and normalized autocovariance functions in subsections \ref{momentmaps}, \ref{vsf}, and \ref{acf}, respectively.
Throughout these subsections, all mock observations were produced in a consistent manner (i.e. consistent orientations and distances).
Later, in subsection~\ref{cloudorientation_clouddistance}, we consider how changing the orientations and distances used to produce mock images alters our results.

To understand the geometric relations between the simulations and mock images used in the first 3 subsections (\ref{momentmaps} - \ref{acf}), it is instructive to reference the panel in the left column of Figure~\ref{fig:sims_tcc03_2sigma} for simulation C, which shows a representative column moment map.
The ray between the observer and the center of the simulation domain, which provides data for the mock observation's central pixel at ($\alpha$, $\delta$) =  ($304^\circ$, $1.5^\circ$), runs parallel to the simulation's z-axis.
The distance along this ray is fixed at 12.4 kpc, the observational estimate for the distance of the SC \citep{Putman2003, Lockman2008, Wakker2008, Fox2016}.

We emphasize that the central pixel is the only pixel where the ray along the line of sight (LOS) perfectly aligns with the simulation's z-axis. For example, the ray used to produce the pixel at ($312^\circ$, $1.5^\circ$) lies in the simulation's x-z plane and there is an 8 degree angle (from $312^\circ-304^\circ$) between the ray and the z-axis. We remind the reader that the wind runs parallel to the simulation domain's x-axis, which is orthogonal to the LOS at the center of the image. Plots are oriented such that the wind always enters from the right-hand side, traveling to the left-hand side.

\subsection{Moment Maps}\label{momentmaps}

\subsubsection{Spatial Distributions}

To determine physical quantities from the observed and mock observed data cubes, we calculate the spectral moments. Spatial distributions (in sky coordinates $\alpha$ and $\delta$) of the resulting moment maps for the observations are shown in the right column of Figure~\ref{fig:observations_smoothed_2sigma}. Here, the Smith Cloud is seen to have it's famous cometary appearance, with the tail of the cloud after the 2.5$\sigma$ clipping being visible around $\left( \alpha, \delta \right) \approx (310^{\circ}, 2.5^{\circ}$). Column densities (first row of Figure~\ref{fig:observations_smoothed_2sigma}) decrease around the edges of the cloud, and the intensity-weighted mean velocity (second row) appears to increase towards the center, with a large majority of the tail travelling at $\overline{v}_{\rm LSR}\leq80$ km/s.

Spatial distributions for the simulations at two points in time, $t=3t_{\rm cc}$ and $t=12t_{\rm cc}$, are shown in Figures~\ref{fig:sims_tcc03_2sigma} and \ref{fig:sims_tcc12_2sigma}, respectively, where each row represents a simulation and each column is one of the moments. As mentioned, the right ascension and declination range in the mock observations are chosen to match the Smith Cloud observations.

In Figure~\ref{fig:sims_tcc03_2sigma}, after only a few cloud-crushing times, simulations A (top row) and B (second row) are nearly identical, largely due to the fact that the only differing simulation parameter is metallicity. Even at this early time, though, a contrast can already be seen in the morphology of simulations C and D, which have the same initial radius. Simulation C lengthens as it becomes entrained within the wind, while simulation D decreases in size as it homogenizes with the background flow. 

At $t/t_{\rm cc}=3$, the \emph{maximum} fraction of the gas (by mass) that originated within the cloud is approximately one for all simulations. This indicates that nearly all of the gas mass present is from the initialized spherical cloud, and minimal mass from the wind has been accreted.

At later times in Figure~\ref{fig:sims_tcc12_2sigma}, it becomes evident that only simulations A and C have grown, with Simulation C doing so at a faster rate than A. We exclude simulation B from this figure as no data is visible; we note that this is due to the 2.5$\sigma$ clipping threshold, and a small fraction of the cloud remains, growing at times later than $12t_{\rm cc}$. Simulation D, however, has been completely destroyed. Each displayed simulation (A and C) show clear signs of the filamentary morphology of growing clouds \citep{Cooper2009, Gronke2018}. 

While Figures~\ref{fig:sims_tcc03_2sigma} and \ref{fig:sims_tcc12_2sigma} only represent the simulations' spatial distributions at two distinct cloud-crushing times (of twelve total in this paper), both at different stages of the cloud-wind evolution (pre- and post-entrainment or near/full destruction), it appears that neither are strikingly similar to the Smith Cloud. However, we caution against only using these two cloud-crushing times as the definitive factor for determining whether a good match to the SC is simulated. The SC appears to extend across a range of about three degrees in declination, most similar to simulations A or B in Figure~\ref{fig:sims_tcc03_2sigma}, though with column densities and tail structures more reminiscent of simulation C. Sigma clipping has some effect on this, as the tails of A and B are nearly cut out, this means that much of the tail of the observed SC is at higher column densities than some of our simulations. Furthermore, the SC shows the lowest velocity dispersions at the very edges of the visible H\MakeUppercase{\romannumeral1} gas structures, and shows a large spread in $\overline{v}_{\rm LSR}$ throughout the cloud, which is more represented by simulation C in Figure~\ref{fig:sims_tcc12_2sigma}.

\subsubsection{Column Density vs. Velocity}\label{m0vsm1}

\begin{figure*}
	\includegraphics[width=\textwidth]{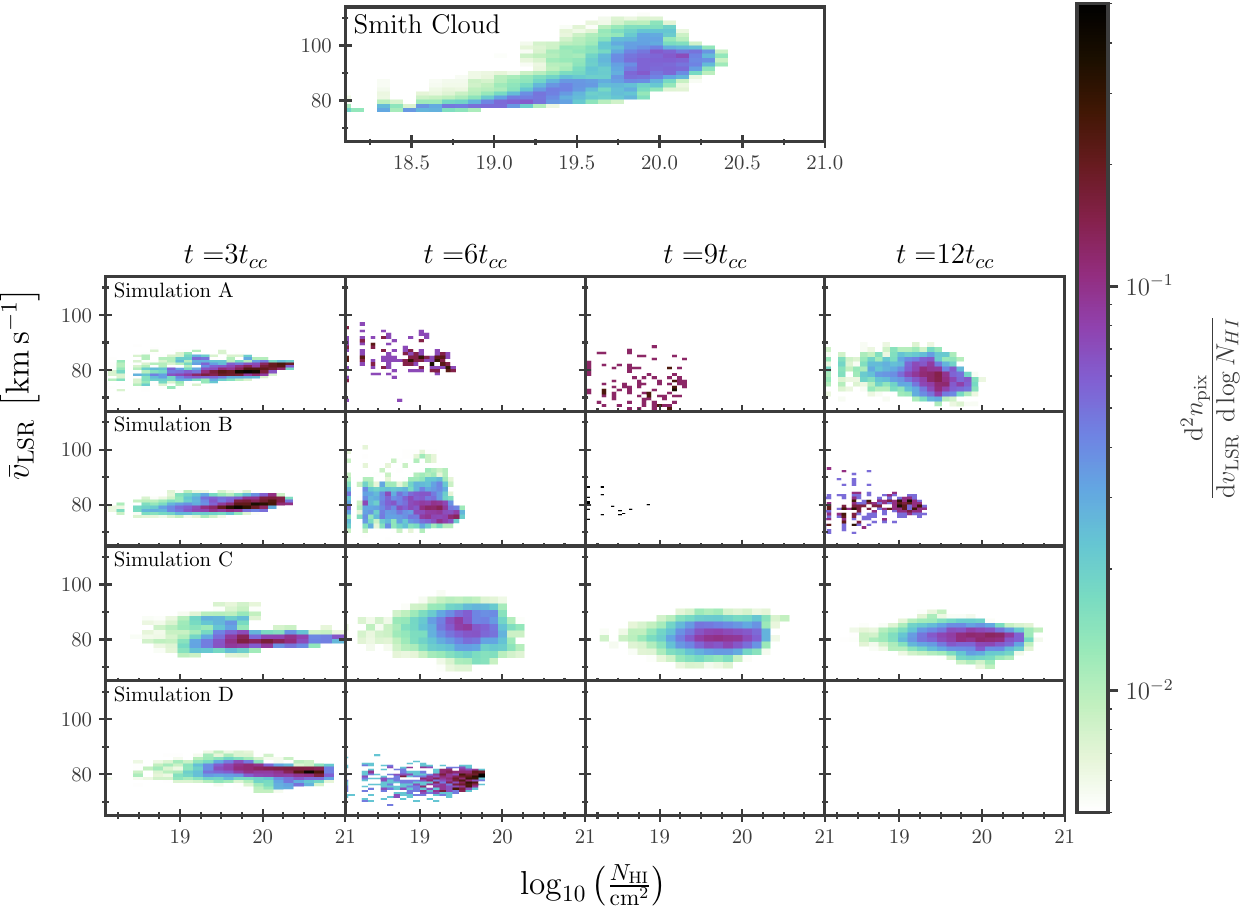}
    \caption{2-D histograms of column density (zeroth moment) and $\overline v_{\rm LSR}$ (first moment) of the Smith Cloud GALFA-HI observations (top panel) and all simulations (lower panels), colored by density of pixels. Rows are organized by simulation, while columns represent different cloud-crushing times. At $3t_{\rm cc}$ (visible in Figure~\ref{fig:sims_tcc03_2sigma}), simulations A and B are still identical, while simulation C becomes entrained in the wind, and simulation D is in the process of being destroyed. By $6t_{\rm cc}$, all simulations are nearly destroyed except for simulation C, and A is seen to be destroyed slightly faster than B. However, simulation A with a metallicity of $Z_{\odot}$ and therefore more efficient cooling, begins to recover, and starts clearly growing by $12t_{\rm cc}$. A few pixels from B are also visible at this time, and we again note that the mass recovers at a later point in time. Simulation D, our adiabatic run modelled after C, is completely destroyed.}
    \label{fig:moment0moment1}
\end{figure*}

While the the qualitative comparisons of the moment maps' spatial morphology distributions is extremely insightful, it is hard to quantify differences between simulations in different stages of evolution.
To make systematic differences easier to see, we examine the joint distributions for pairs of moments (column density, intensity-weighted LOS velocity, and intensity-weighted mean velocity dispersion) that are measured from a given mock observation. 
With other pairs being discussed in subsequent subsections (see \ref{m0vsm2} and \ref{m1vsm2}), we begin by examining the $N_{\rm H\MakeUppercase{\romannumeral1}}$ versus $\overline v_{\rm LSR}$ joint distribution in Figure~\ref{fig:moment0moment1}.

Colored by the pixel density in each bin, the top panel of Figure~\ref{fig:moment0moment1} shows the joint moment distribution for the Smith Cloud (observational values), which demonstrates a clear trend and appears to be split into two populations: one with distinctly higher velocities and column densities, and one with lower velocities and column densities. 

The lower panels of Figure~\ref{fig:moment0moment1}  show data measured from the simulated clouds at different times, with rows organized as simulations and columns as specific cloud-crushing times ($t/t_{\rm cc}=3, 6, 9, 12$). At $3t_{\rm cc}$ (first column), the simulations' LOS velocity measurements seem to span a lower range than the Smith Cloud, especially at the highest column densities. By $6t_{\rm cc}$ (second column), the reduced pixel distribution for simulations A, B, and D suggests that the clouds are possibly in the process of being destroyed.  At this time for simulation C, however, the LOS velocity measurements span a wider range of 25 km/s, which is comparable to the range of the Smith Cloud. 

At $t=9-12t_{\rm cc}$ (third and fourth columns), the increasing amount of data illustrates that simulation A starts to rapidly grow, while simulation D is completely destroyed. Simulation C appears to narrow in velocity range, and the column density increases. We note that while simulation B also appears to be destroyed, this is a  result of the 2.5$\sigma$ clipping. The colder gas is nearly homogenized with the hotter gas, but (similar to simulation A), it begins to grow at a later time, which is indicated by the reappearance of data at $12t_{\rm cc}$.

However, while our mock observations show a similar spread of column densities, no mock observation appears to match the Smith Cloud's correlation between the column density and average velocity: instead, mock observed clouds maintain a relatively flat average velocity, particularly at late time.

As mentioned in Section~\ref{momentmaps}, our sigma clipping level of $2.5\sigma$ removes both noise and part of the clouds' tails, especially in the lower $N_{\rm H\MakeUppercase{\romannumeral1}}$ simulations, A and B. This includes more diffuse H\MakeUppercase{\romannumeral1} gas, which we observe in a few small areas of the tails to have higher velocities and velocity dispersions than represented by Figures~\ref{fig:observations_smoothed_2sigma} and ~\ref{fig:sims_tcc03_2sigma}. However, we find that less stringent sigma clipping (e.g., $1-2\sigma$) does not significantly change the distributions displayed by Figure~\ref{fig:moment0moment1}, or the  joint moment distribution functions discussed in the next subsections. Reducing the restrictive nature of the sigma clipping shows the same primary distributions, albeit with more scatter (of low pixel densities, $\sim0.01$), especially at lower column densities. This is expected due to the nature of the clipping.

\subsubsection{Column Density vs. Velocity Dispersion}\label{m0vsm2}

\begin{figure*}
	\includegraphics[width=\textwidth]{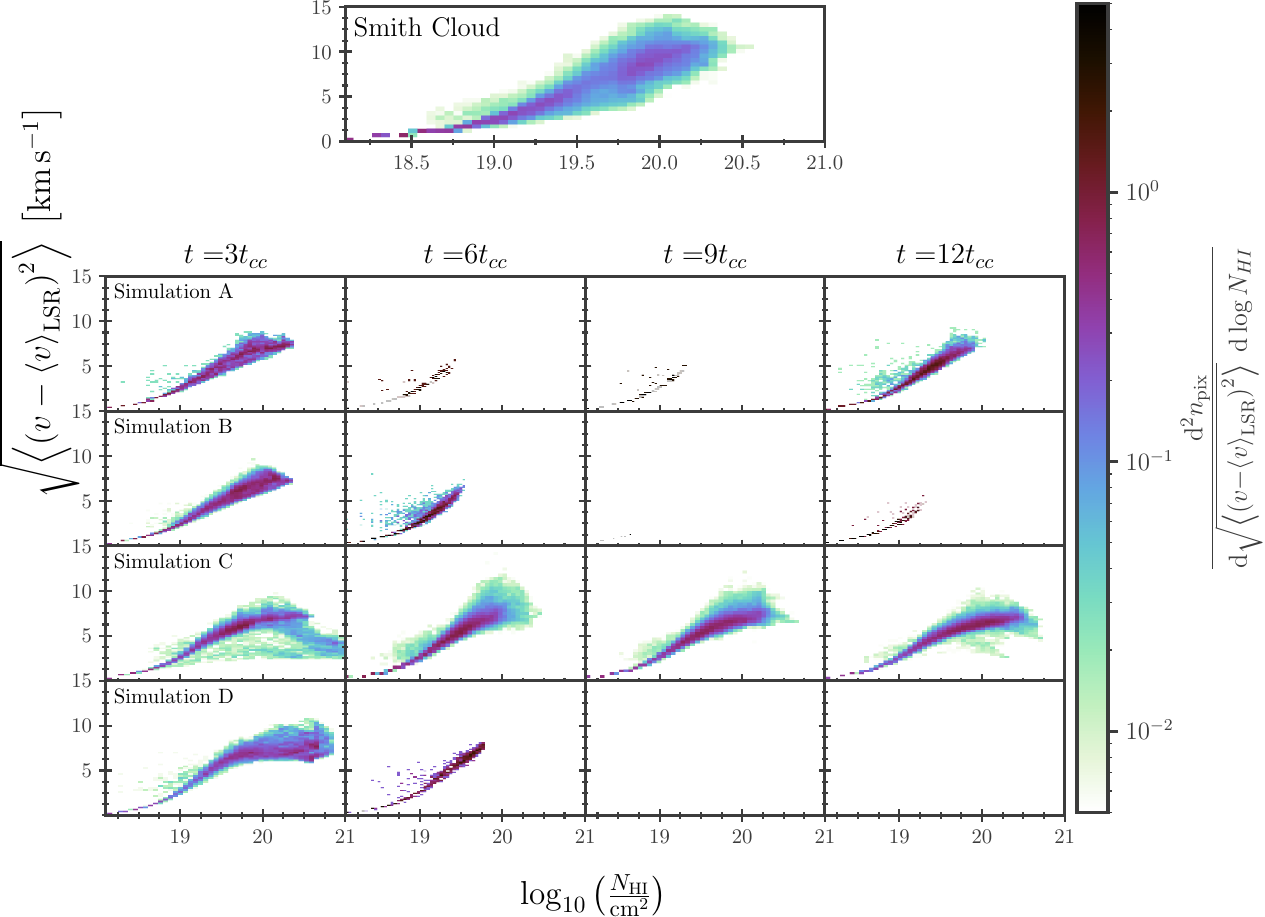}
    \caption{2-D histograms of column density ($N_{\rm H\MakeUppercase{\romannumeral1}}$; zeroth moment) and velocity differences in the cloud ($\sqrt{\left< (v-\left<v\right>_{\rm LSR})^2 \right>}$; second moment) in the style of Figure~\ref{fig:moment0moment1}. At later times, surviving simulations appear similar to observations, with a cometary shape trailing off to lower velocity differences and column densities. Simulation C, however, is the only simulation that reaches column densities similar to those of the observed Smith Cloud beyond $3t_{\rm cc}$. }
    \label{fig:moment0moment2}
\end{figure*}

Figure~\ref{fig:moment0moment2} is similar to Figure~\ref{fig:moment0moment1}, but  instead shows the joint distribution of column density (zeroth moment) and LOS velocity dispersion (second moment). 

The distribution measured from the SC observation maintains a similar cometary morphology to that found in Figure~\ref{fig:moment0moment1}, indicating that lower column density gas in the Smith Cloud also has lower velocity dispersions, while the higher column density gas maintains larger velocity dispersions. In addition, the observations form a bimodal distribution: the higher column density distribution appears to be $N_{\rm H\MakeUppercase{\romannumeral1}}>10^{19.75} \rm cm^{-2}$, while the lower column density distribution is $N_{\rm H\MakeUppercase{\romannumeral1}}<10^{19.5} \rm cm^{-2}$.

In contrast to the the $N_{\rm H\MakeUppercase{\romannumeral1}}$ versus $\overline v_{\rm LSR}$ plots, the distributions  in Figure~\ref{fig:moment0moment2} measured from the mock observations are qualitatively similar to that of the SC. Pixels with larger column density generally have larger LOS velocity dispersions, particularly at late times. 
There are 2 exceptions to this trend, both at $t=3t_{\rm cc}$.
First, Simulation C appears to have a small distribution of pixels, separate from the main distribution, at lower velocity dispersion and higher column densities, which disappears at later times.
Second, in Simulation D (at this same time), pixels with $N_{\rm H\MakeUppercase{\romannumeral1}} > 10^{19.75} \; \rm cm^{-2}$ have roughly constant velocity dispersions, rather than increasing. 

\subsubsection{Velocity vs. Velocity Dispersion}\label{m1vsm2}

\begin{figure*}
	\includegraphics[width=\textwidth]{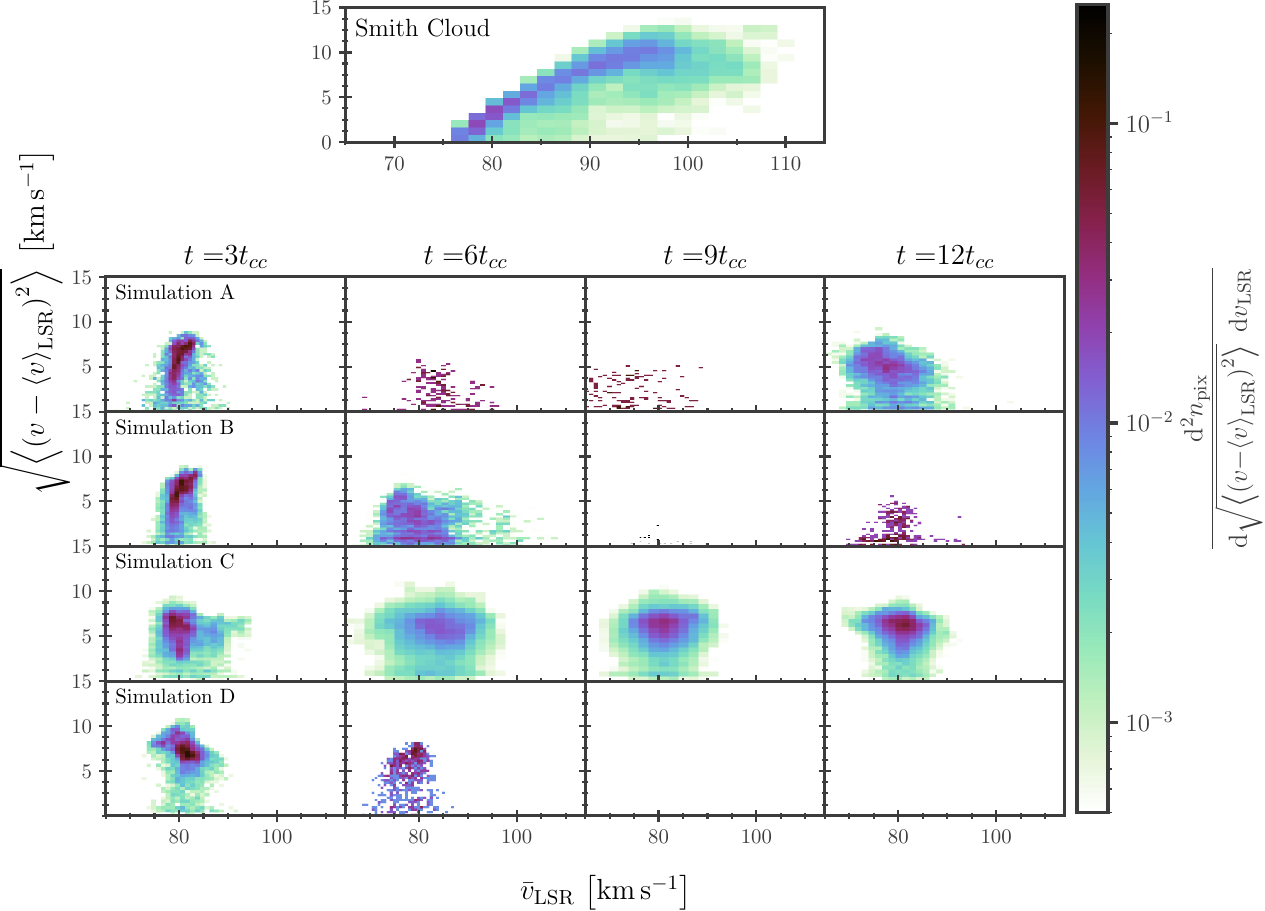}
    \caption{2-D histograms of $\overline v_{\rm LSR}$ (first moment) and velocity dispersion through the cloud ($\sqrt{\left< (v-\left<v\right>_{\rm LSR})^2 \right>}$; second moment) in the style of Figure~\ref{fig:moment0moment1}. No simulation at any point in time appears to match the morphology of the Smith Cloud observations in this parameter space, but the $6t_{\rm cc}$ column clearly shows how this distribution changes as clouds are either entrained in the wind (simulation C, third row) or homogenized (simulation D, and nearly simulations A and B).}
    \label{fig:moment1moment2}
\end{figure*}

Figure~\ref{fig:moment1moment2} displays the distribution of intensity-weighted LOS velocity (first moment) vs. velocity dispersion (second moment) across the cloud. Similarly to Figure~\ref{fig:moment0moment1}, the observations show a quite different behavior than the simulations, whether actively growing or being destroyed.  

The shape of the SC's distribution contrasts the distributions for the simulations at each point in time. Instead of having a central concentration around a specific velocity and velocity dispersion, the  SC's observed distribution retains the cometary (or arc-like) morphology, similar to what is seen in the other parameter spaces. Much of the gas is concentrated at lower velocities ($\overline{v}_{\rm LSR}\approx 77-78$ km/s) and velocity dispersions, and the velocity dispersion increases with increasing velocity.

In the simulations, however, the $\overline{v}_{\rm LSR}$ distribution in the simulations is narrow, rarely reaching above 90 km/s, while the SC shows velocities up to 100+ km/s. The simulations also lack the clear correlation between LOS velocity and velocity dispersion seen in the observations, and there is not a bimodal distribution. 

For example, even as early as $3t_{\rm cc}$ (see Figure~\ref{fig:sims_tcc03_2sigma}), the mock observation of simulation C has a tail with $\overline{v}_{\rm LSR}$ up to $\sim$ 95 km/s, a velocity higher than the bulk of the cloud. Simulation A experiences a similar behavior, but perhaps serves to show that clouds may overcome such a velocity decline to still grow. In Figure~\ref{fig:moment1moment2}, while it is nearly destroyed around $6t_{\rm cc}$ and $9t_{\rm cc}$, the velocity and velocity dispersion both fall, then increase as the cloud is entrained and begins growing by $12t_{\rm cc}$. During this growing stage, simulation A more closely resembles the joint distribution of simulation C. 

 We further discuss possible causes behind the discrepancy in simulations and observations in Section~\ref{discussion}.

\subsection{Projected Velocity Structure Function}\label{vsf}

\begin{figure*}
	\includegraphics[width=\textwidth]{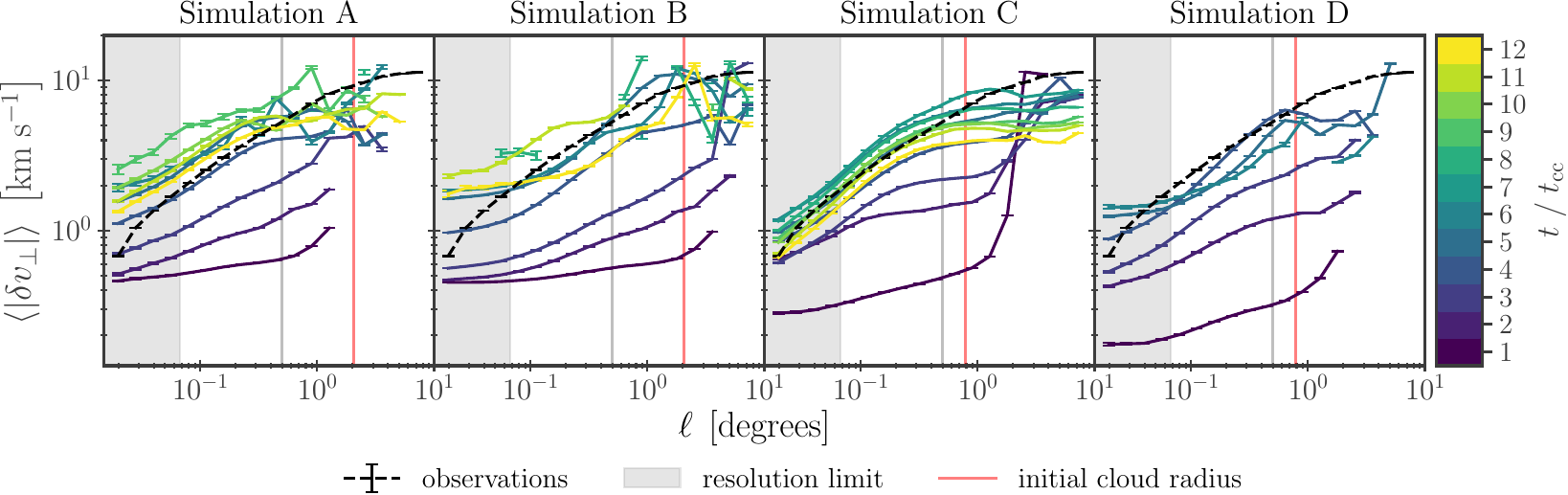}
    \caption{Projected first-order velocity structure functions for each simulation (shown in each column) colored by cloud-crushing time up to $12t_{\rm cc}$. Values measured from the Smith Cloud observations are represented by dashed black lines. We require at least 100 pairs of points in a bin for it to be plotted, and errors are the normalized standard error of the data in that bin. Red vertical lines show the simulated cloud's initial angular size on the sky, while the grey vertical line serves as a point of reference for $\ell = 0.5^{\circ}$. The shaded grey region for $\ell \leq 0.67^{\circ}$ denotes the region below the GALFA-HI resolution limit.}
    \label{fig:vsf}
\end{figure*}

To characterize the turbulence within the clouds, we employ the projected first-order velocity structure function (VSF; \citealt{Frisch1995}). This method characterizes the magnitude of the velocity difference as a function of distance $\ell$, computed using $\overline{v}_{\rm LSR}$. 
Structure functions are useful because theories of turbulence make predictions about how differences in the 3D velocity field vary with $\ell$. 
For example, idealized Kolmogorov turbulence (isotropic, homogeneous, subsonic turbulence in an incompressible fluid) predicts that the velocity scales as $\ell^{1/3}$ up to the length-scale at which the turbulence is driven, where the slope changes \citep{Ossenkopf2002, Federrath2013, Chira2019, Li2020a, Ha2021, Gronke2022, Hu2022, Li2023, Abruzzo2024}. However, we note from both \citet{Li2020a} and \citet{Mohapatra2022} that a projected VSF may look somewhat different as a result of the projection effect.

Following \citet{Abruzzo2024}, we compute the VSF by defining $|\delta v_\perp|$ as the magnitude of the velocity difference between a pair of randomly-selected points $i$ and $j$, with the first-order velocity structure function, $\left< |\delta v_\perp| \right>(\ell)$, being the average value of $|\delta v_\perp|$ for all pairs of points $\left(i, j \right)$ from the first moment map (excluding those where $N_{\rm HI}=0$) separated by a distance of $\ell$. We note here that the computed distance is the Euclidean `pixel' distance using the small-angle approximation and therefore does not account for curvature, though this is unlikely to significantly affect the scales we analyze, and any biases from this approximation would be the same in both the Smith Cloud observations and our mock observations.

Projected first-order velocity structure functions for the observations and all simulations analyzed here are shown in Figure~\ref{fig:vsf}. Each column represents the respective simulation, while each output time distinguished by line color. Observations are plotted in each panel as a dashed black line. We require a minimum of 100 pairs of points to be plotted, although this is only relevant in simulations where the clouds are nearly destroyed. Red vertical lines on each panel show the cloud's initial radius, and therefore the initial limit of $\ell$. We also computed second-order velocity structure functions, $\left< |\delta v_\perp|^2 \right>(\ell)$, but these are qualitatively similar to the first-order structure functions, and therefore we do not show them here.

Despite the differences across the simulated cloud parameters, especially simulation D having no cooling,  it is unsurprising that the projected VSFs are so similar at early times in Figure~\ref{fig:vsf}. Initially, the cloud has little-to-no turbulence as a nearly uniform sphere at rest. At these early times ($t/t_{\rm cc} < 4$), the clouds all experience an increase in the VSF as turbulence is driven by shear at the cloud-wind boundary, as detailed in \citet{Abruzzo2024}. The VSF continues to grow until the cloud is entrained in the wind and the overall velocity difference driving the shear flow drops. This can be seen most clearly in the long-survival run (simulation C).  We discuss physical and observational impacts on the VSF in more detail in Section~\ref{discussion_VSF}.

Simulation A is a reasonable match to the SC starting around $5t_{\rm cc}$, though it experiences deviations at higher $\ell$ as the cloud's intensity falls below our observational limits. 
Simulation B is similar, though it is less of a match to observations: the VSF on larger scales becomes more varied, or more noisy, than observations (particularly as the cloud undergoes near-destruction) largely due to the few remaining cloud clumps at large $\ell$.

Simulation C sees the fewest changes in its VSF after $3t_{\rm cc}$, with increasing time only slightly lowering the magnitude of $\left< |\delta v_\perp| \right>$. The VSF flattens at higher separations (around $\ell\sim0.15^{\circ}$), but is, overall, remarkably similar to observations below the degree scale. Because simulation D does not grow, the early cloud-crushing times are similar to C, but it does not ultimately survive.

\subsection{Normalized Autocovariance Function (ACF)}\label{acf}

\begin{figure*}
	\includegraphics[width=\textwidth]{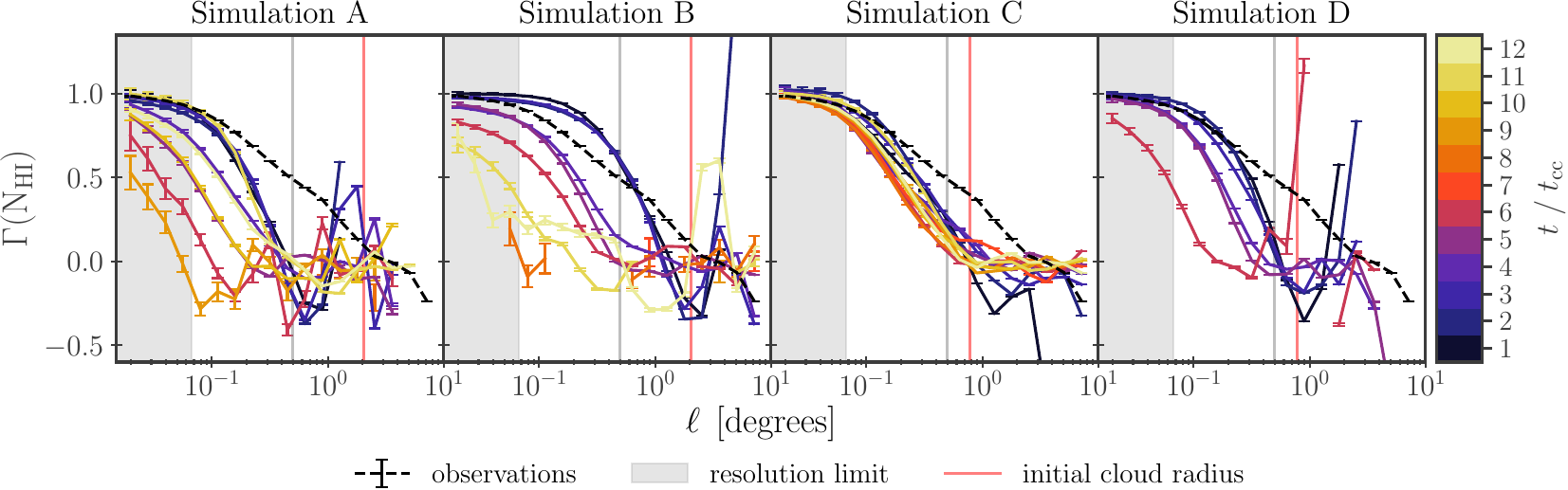}
    \caption{Normalized autocovariance function of column density for each simulation (column) colored by cloud-crushing time up to $12t_{\rm cc}$, similar to Figure~\ref{fig:vsf}. Observations are represented by dashed black lines. Errors are the normalized standard error of the data in each bin, where we require at least 100 pairs of points in each bin. Red and grey vertical lines and shaded grey region are as Figure~\ref{fig:vsf}. A clear contrast exists between simulations that are being destroyed or growing, as the shrinkage of destroyed clouds is evident.}
    \label{fig:2pc}
\end{figure*}

Just as the VSF provides insight into measures of velocity differences and turbulence across the cloud, we use the normalized autocovariance function (ACF) to do the same for column density. 

The normalized autocovariance function is defined as

\begin{equation}\label{acf_eqn}
    \Gamma(N_{\mathrm{HI}}) = \frac{\sum \left(N_{\mathrm{HI}}(i)- \left< N_{\mathrm{HI}} \right> \right)\left( N_{\mathrm{HI}}(j)- \left< N_{\mathrm{HI}} \right> \right)}{\sum \left(  N_{\mathrm{HI}}(i)-\left< N_{\mathrm{HI}}\right> \right)^2},
\end{equation}
where the summation, similar to the velocity structure function, is over all randomly-selected pairs of pixels $\left(i, j\right)$ with separations between separation $\ell$ and $\ell+\Delta\ell$. Ultimately, this function measures the correlations in the deviation from the mean intensity. 

Using this definition, a value of 1 indicates a perfect match in column density between the pair $(i, j)$ (i.e., $N_{\rm H\MakeUppercase{\romannumeral1},i}=N_{\rm H\MakeUppercase{\romannumeral1},j}$), and we would expect the function to reach zero around the cloud's radius, indicating the size of the cloud. To avoid a dependence on the size of the image, we do not include zeros in this calculation.

Figure~\ref{fig:2pc} shows the ACF for the same set of simulations, again colored by cloud-crushing time. Unlike the VSF, it does not appear that any of our simulations are a strong match to the observations. However, we are able to identify possible physical drivers of this measure. 

All simulations start around $\Gamma(N_{\rm H\MakeUppercase{\romannumeral1}}) \approx 1$ on small scales for early cloud-crushing times, similar to the Smith Cloud. Simulations A, C, and D match the observed slope up to $\ell=0.1^{\circ}$ at early cloud-crushing times, beyond which most simulations show a steeper slope. In addition, while simulation C shows slight variations in the function beyond $\ell \approx 0.8^{\circ}$, it largely flattens to values around 0 (implying no correlation on large scales), which also corresponds to the cloud's initial size as mentioned above.

We note here that adding noise to the simulations and introducing sigma clipping has a significant effect on the function's appearance, explaining the noisy behavior seen in several panels of Figure~\ref{fig:2pc}. This is particularly pronounced at larger $\ell$ values and for clouds that do not survive.
For high-intensity simulations, such as C, the point at which the function reaches 0 may serve as a possible indication of the cloud depth. Indeed, simulation C demonstrates behavior that is both steady in time and in reasonable agreement with the observed ACF on the sub-degree scale but not beyond. This seems consistent with the moment 0 maps in Figure~\ref{fig:observations_smoothed_2sigma} and Figure~\ref{fig:sims_tcc12_2sigma}.

\subsection{Dependence on Cloud Orientation and Distance}\label{cloudorientation_clouddistance}

Halo clouds have a wide range of distances, from relatively close to the Milky Way's disk to nearly 100 kpc for the Magellanic Stream \citep{Putman2012}. In addition, the viewing angle (relative to the cloud motion) for clouds is rather difficult to constrain. To determine how variations in both viewing angle with respect to the observer and distance can affect the statistics we have presented, we vary both properties for simulation C and investigate how (and if) results differ. Simulation C is chosen as it is the cloud that most clearly grows and is arguably the closest match to the observations, although the trends we find extend to other clouds.

\begin{figure}
    \centering
    \includegraphics[width=\columnwidth]{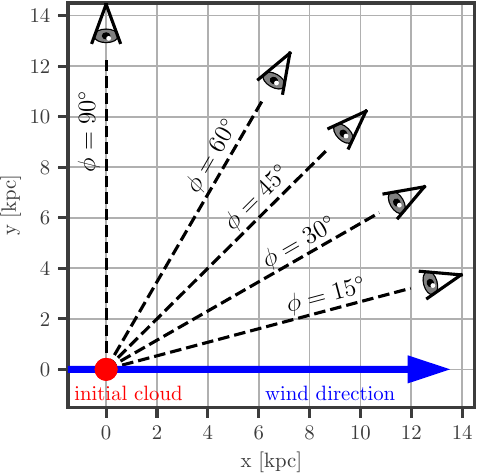}
    \caption{Definition of our viewing angle $\phi$. Black dashed lines represent the ray directions for each angle or the line of sight, and the red sphere represents the initial spherical cloud of our simulations. The blue arrow along the +x-axis shows the direction of the hot wind. $\phi=90^{\circ}$ corresponds to a viewing angle transverse to the wind.}
    \label{fig:angle_projection}
\end{figure}

\subsubsection{Varying Cloud Orientation}

While all previous results have used a uniform viewing angle in spherical coordinates of $(r, \theta, \phi) = (d, 90^{\circ}, 90^{\circ}$), where $d$ is the cloud distance, we now keep $\theta$ constant at $90^{\circ}$ and vary $\phi$ to $15^{\circ}$, $30^{\circ}$, $45^{\circ}$, $60^{\circ}$, and the previously-shown $90^{\circ}$. The orientation of these viewing angles is visible in Figure~\ref{fig:angle_projection}. The red sphere represents the initial cloud, while the wind travels along the +x-axis, denoted by the blue arrow. The Smith Cloud is predicted by \citet{Lockman2008} to be at a viewing angle of $45^\circ\pm10^\circ$.

\begin{figure*}
    \centering
    \includegraphics[width=\textwidth]{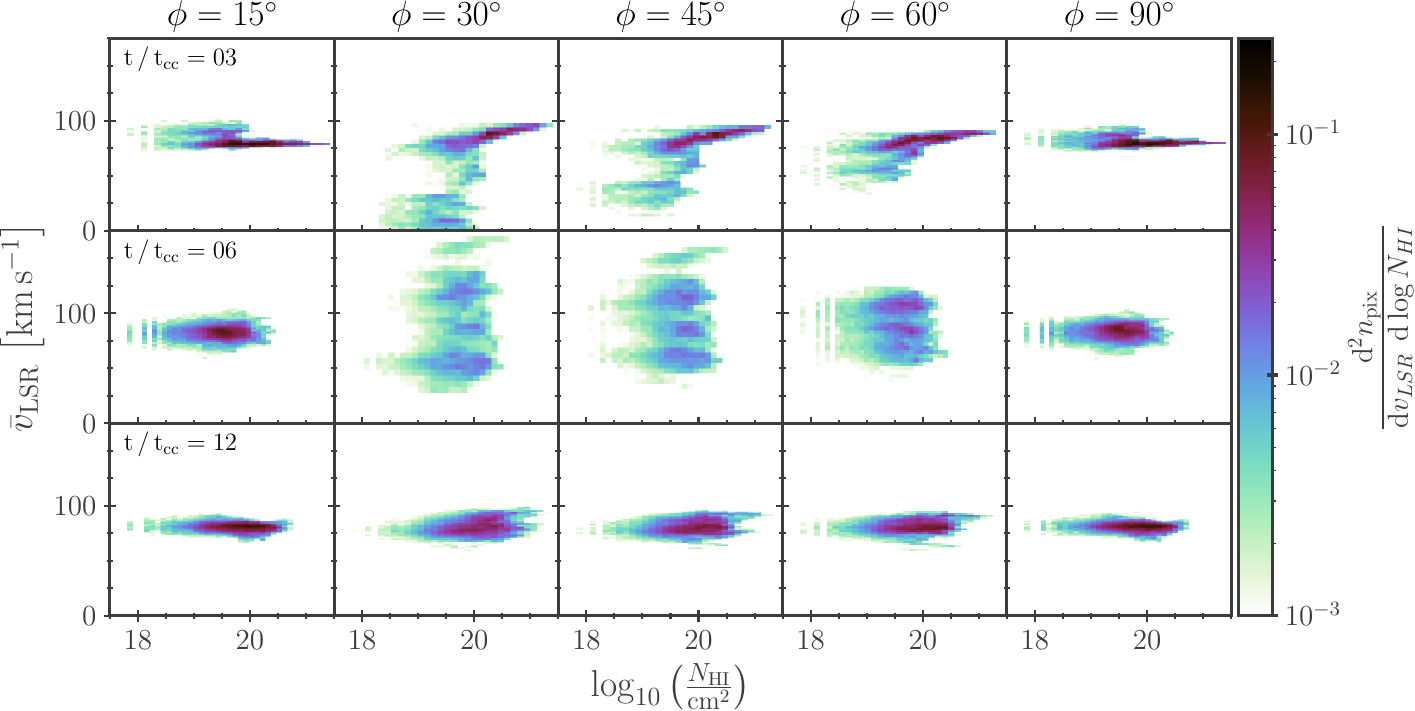}
    \caption{2-D histograms of column density versus intensity-weighted average velocity, similar to Figure~\ref{fig:moment0moment1}, now only for simulation C, and for varying viewing angles relative to the observer; we also note the extended $\overline{v}_{\rm LSR}$ axis compared to Figure~\ref{fig:moment0moment1}. Rows represent cloud-crushing times of $3t_{\rm cc}$ (top), $6t_{\rm cc}$ (middle), and $12t_{\rm cc}$ (bottom). Columns are organized by the angle $\phi$ in spherical coordinates, where we keep $\theta$ constant at $90^{\circ}$, as before, and vary $\phi$ from $15^{\circ}$ (first column) up to $90^{\circ}$ (last column). The orientation of the cloud appears to only subtly affect column densities at $12t_{\rm cc}$, but significantly affects $\overline{v}_{\rm LSR}$, with $\phi=30-60^{\circ}$ showing wider velocity ranges of nearly an order of magnitude. This is lessened at the later cloud-crushing time. }
    \label{fig:m0m1_angles_3612}
\end{figure*}

Figure~\ref{fig:m0m1_angles_3612} is similar to Figure~\ref{fig:moment0moment1}, displaying 2-D histograms of column density versus intensity-weighted velocity. However, we now only show results for simulation C, calculated at cloud-crushing times of 3 (top row), 6 (middle row), and 12 (bottom row), and for the specified angles of $\phi$ (different columns). Despite similarities to Figure~\ref{fig:moment0moment1}, we also note the substantially increased dynamic range of the velocity axis (and remind readers that, in the observational case, velocities below about 65 km/s are hard to interpret due to possible confusion with the Milky Way). 

For the intermediate angles of $\phi=30-60^{\circ}$, the range of intensity-weighted velocities increases significantly compared to the previous results from $\phi=90^{\circ}$, which have only minute differences from $\phi=15^{\circ}$. However, this depends sensitively on the stage of the cloud-wind interaction. When the cloud is early in its interaction with the hot wind, at $3t_{\rm cc}$, $\overline{v}_{\rm LSR}$ for $\phi=30^{\circ}$ covers the 0-100 km/s range, the lower limit of which then increases with increasing $\phi$, before finally decreasing to the previously observed range of $\approx75-100$ km/s at $\phi=90^{\circ}$. These LOS velocities then become even more widely distributed at $6t_{\rm cc}$ before collapsing into a narrower distribution at $12t_{\rm cc}$, by which time the cloud has been growing for some time and is largely entrained in the wind. While angles of $\phi=30-60^{\circ}$ at this time still show slightly larger variations in $\overline{v}_{\rm LSR}$ than for $\phi=15^{\circ}$ and $\phi=90^{\circ}$, the effect is considerably less than at earlier cloud-crushing times.  
As a result, it appears the physical perspective of  $\phi=90^{\circ}$ fails to capture the bulk velocity along the length of the cloud, while lower angles become dominated by it.

We note that the shape of the distributions at $t = 3t_{\rm cc}$ and $\phi=30-60^{\circ}$ appears to be a much better match to the Smith Cloud than at $\phi=90^{\circ}$ in Figure~\ref{fig:moment0moment1}. In particular, higher column density material for $\phi=30-60^\circ$ also has higher velocities, and the velocity matches the observational range of about 75-100 km/s. 

Although we do not show it here, the velocity dispersion varies with viewing angle similarly to the LOS velocity, as expected. However, despite the different orientations, we still do not see a good match for the Smith Cloud's joint distribution of LOS velocity and velocity dispersion (Figure~\ref{fig:moment1moment2}).

\begin{figure*}
    \centering
    \includegraphics[width=\textwidth]{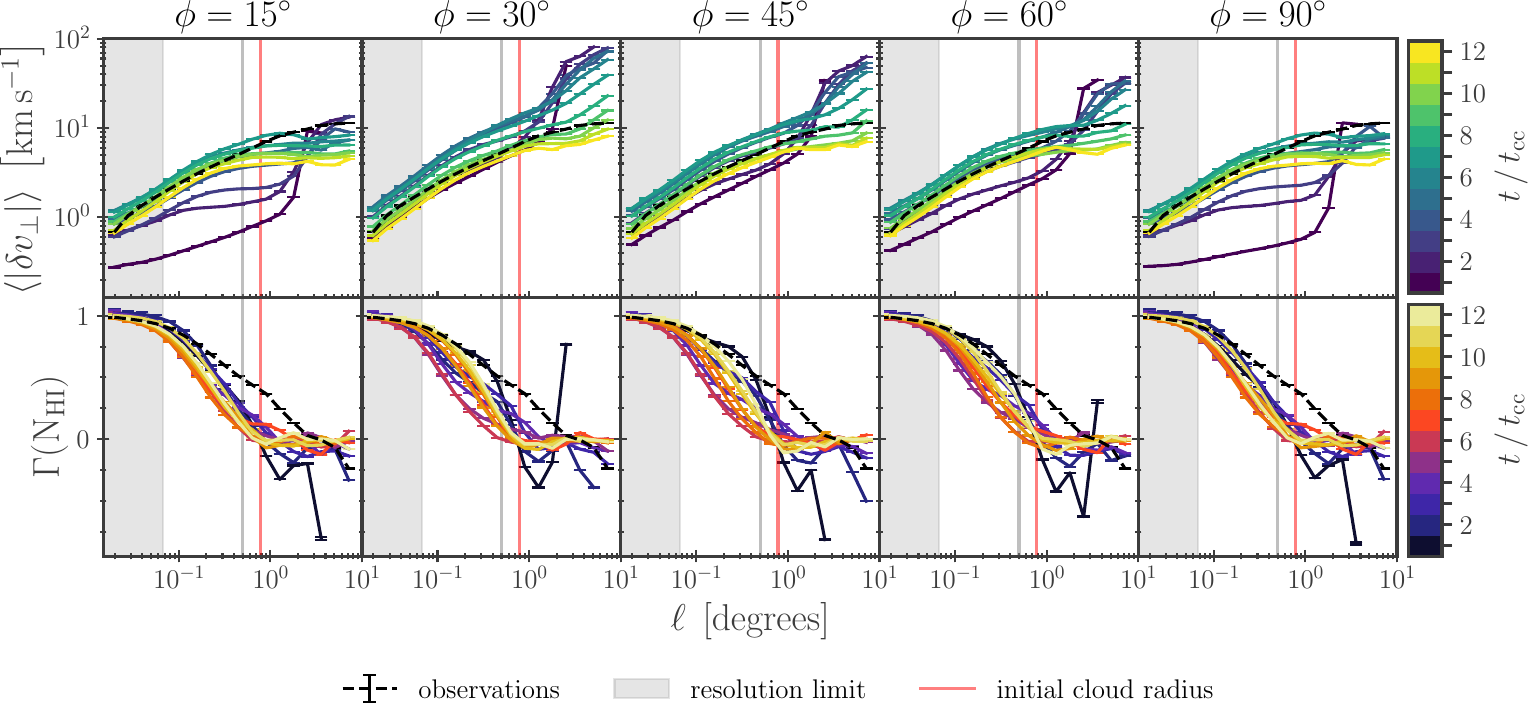}
    \caption{Projected velocity structure function (top row) and normalized autocovariance function of column density (bottom row) of simulation C, for varying cloud orientations ($\phi=15-90^{\circ}$). Lines are colored by cloud-crushing time, as Figures~\ref{fig:vsf} \& \ref{fig:2pc}, respectively. The angle of cloud orientation appears to significantly affect the VSF, with $\phi=30-60^{\circ}$ showing larger values of $\langle | \delta v_\perp | \rangle$ at earlier cloud-crushing times across all $\ell$. In addition, instead of the VSF flattening at larger $\ell$, $\phi=30-60^{\circ}$ result in an increase of $\langle | \delta v_\perp | \rangle$. However, the autocovariance function does not seem to be significantly affected, with only slight variations at low cloud-crushing times, similar to the conclusion from Figure~\ref{fig:m0m1_angles_3612}. }
    \label{fig:VSF_ACF_angles}
\end{figure*}

Finally, the effect of viewing angle on the velocity structure and autocovariance functions of simulation C, at cloud-crushing times up to 12, can be found in Figure~\ref{fig:VSF_ACF_angles}. Of the two functions, the VSF changes significantly more with $\phi$ than $\Gamma (N_{\rm HI})$. 

The shape of the VSF is significantly different for $\phi=30-60^{\circ}$, especially at both early times \textit{and} large $\ell$. 

For these angles ($\ell > 2^{\circ}$), where the VSF previously flattened to a near-constant value at cloud-crushing times greater than three, it now continues to increase, stretching up to values an order of magnitude higher than $\phi=0^{\circ}$ or $90^{\circ}$. As before, $\phi=30^{\circ}$ shows the biggest shift, with VSF values decreasing with increasing $\phi$ until the VSF again flattens at large $\ell$ for $\phi=90^{\circ}$. We find that the intermediate angles of $\phi$ ($30-60^{\circ}$), particularly at late times, show a better match to the projected VSF of the Smith Cloud.

Larger scales of $\ell>1^{\circ}$ are more sensitive to the bulk velocity along the cloud. This is evident across simulations A-D, where simulations undergoing (near) homogenization with the hot wind show larger variation in their VSF, or simply do not reach $\ell$ larger than that of the cloud's initialized size. This also explains why the velocity structure functions are more sensitive to viewing angle but less so to resolution. The bulk motions of the cloud are generally resolved, but changing the line-of-sight with respect to the cloud and wind allows for a different perspective of the bulk motions driven by the wind in the $+\hat{x}$-direction.

Despite the dependence of the VSF on cloud orientation, the autocovariance function of column density, $\Gamma(N_{\rm H\MakeUppercase{\romannumeral1}})$, shows very little quantifiable dependence (bottom row of Figure~\ref{fig:VSF_ACF_angles}. Again, we note that the results for angles of $\phi=15^{\circ}$ and $\phi=90^{\circ}$ are nearly identical. This is unsurprisingly given that the primary impact of viewing angles appears to be the impact of the bulk velocity along the cloud.

\subsubsection{Varying Cloud Distance}

\begin{figure}
    \centering
    \includegraphics[width=\columnwidth]{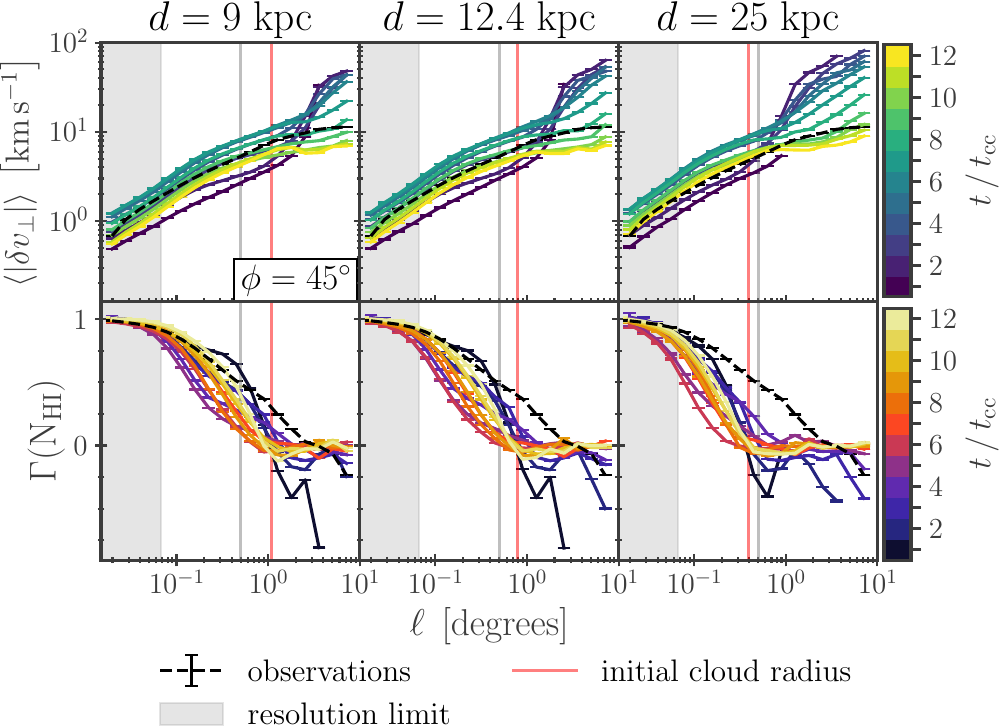}
    \caption{Projected velocity structure function (top row) and normalized autocovariance function of column density (bottom row) for simulation C, but now varying cloud distances (columns) at the same angle of $\phi=45^{\circ}$. The middle column represents the observed distance of the Smith Cloud, while the first and second columns are 75\% and 200\% of this distance, respectively. Red vertical lines show the simulated cloud's initial angular size on the sky, while the grey vertical line serves as a point of reference for $\ell = 0.5^{\circ}$. The shaded grey region for $\ell \leq 0.67^{\circ}$ denotes the region below the GALFA-HI resolution limit. Both functions appear to be shifted towards smaller $\ell$ as distance (and initial angular size) increase.}
    \label{fig:VSF_ACF_distance}
\end{figure}

Finally, we turn to cloud distance. Because the joint distributions (2-D histograms) across each set of moments are not significantly impacted by increasing or decreasing the cloud distance, we do not show them here. Instead, Figure~\ref{fig:VSF_ACF_distance} shows the VSF (top row) and ACF (bottom row) with a constant viewing angle of $\phi=45^{\circ}$, but changing cloud distance. $\phi=45^{\circ}$ is chosen here to highlight the variation in both functions with distance, as the effect is considerably more subtle at our initial $\phi=90^{\circ}$.  The middle column represents the predicted distance of the Smith Cloud from previous literature, $d = 12.4$ kpc, and we scale this distance by $\sim75\%$ to 9 kpc (first column), and $\sim200\%$ to 25 kpc (last column).

Distance does not significantly impact the overall shape of these functions, although slight effects may be seen, but rather it changes the angles, stretching or compressing the distribution with respect to $\ell$. At the observed distance of 12.4 kpc, the VSFs appear to change their slope and increase right at $\ell = 2^{\circ}$. 
This behaviour is expected given the relation between angular size and physical distance. Agreement with observations (particularly for the ACF) improves with the closest distance ($d=9$ kpc).

\section{Discussion}\label{discussion}

Following the presentation of our Smith Cloud GALFA-HI observations and simulated mock observations, we now consider our results in a broader context and compare to previous literature. We first consider whether our simulated (mock observed) clouds match the observed Smith Cloud (\ref{reproduce}), then revisit the physics involved in the velocity structure function (\ref{discussion_VSF}), before turning to possible implications for observers (\ref{discussion_implications}), and limitations of this study (\ref{caveats}).

\subsection{Can Simulations Reproduce Observations?} \label{reproduce} 

We have demonstrated that it is possible to carry out a detailed comparison between mock observations of cloud/wind interactions and radio data cubes, and several individual metrics can be reproduced (or nearly reproduced). We find that simulations A and C are the closest overall matches to the Smith Cloud. We further highlight the relevance of TRML entrainment, how metrics shown change with different simulation properties, their physical drivers, and whether they may serve as a diagnostic for cloud survival or destruction. Given the position of the Smith Cloud in the lower halo of the MW, we emphasize that a similar simulated cloud likely needs to be stable against destruction (homogenization with the halo).

\subsubsection{Relevance of TRML Entrainment to HVCs}\label{discussion_TRML}

This paper was designed to study HVCs through TRML entrainment models (and vice versa) with recent high-resolution observations and simulations. While it was previously theorized that TRML entrainment is relevant for HVCs, we have now placed this on firmer observational footing. In particular, given the cloud initial conditions, we find that our best match simulation is the one clearly showing TRML entrainment.

Section~\ref{simulationinfo} outlines the model that simulated initial clouds follow. In short, Equation~\ref{clouddensity} encodes the basic arguments about cloud properties that result in \textit{highly} constrained initial conditions if we wish to match what has been observed of the Smith Cloud. In fact, because we have observational measurements for the mass $M_{\rm cl}$ and metallicity $Z_{\rm cl}$, the \textit{only} free parameter is thermal pressure (outside of initial cloud velocity and shape). As a result, significantly lowering the thermal pressure is likely the only way that cooling does not significantly alter cloud evolution. 

Indeed, the importance of radiative cooling is evident in simulation B. Although the only difference between A and B is the lower metallicity, simulation B lies on the threshold of survival, and appears to be destroyed if data below observational limits are ignored. The detailed physics of turbulent radiative mixing layers is then crucial to the cloud's fate. 

\subsubsection{Column Density, Velocity, and Velocity Dispersion} \label{reproduce_moments}

We are able to replicate the Smith Cloud's distribution of H\MakeUppercase{\romannumeral1} column densities, derived from the zeroth spectral moment, in simulation C (growing cloud). We note that this is feasible due to the relationship between column density and initialized cloud properties, explained in Section~\ref{simulationinfo} ($R_{\rm cl} \propto \rho_{\rm cl}^{-1/3}$, $R_{\rm cl} \propto n_{\rm cl}^{-1/3}$,  $N_{\rm cl} \propto n_{\rm cl}^{2/3}$). As a result, after initial runs with simulations A and B, it is fairly simple to adjust initial conditions such as thermal pressure $p/k_B$ and radius $R_{\rm cl}$ in order to approximately produce the desired values of $N_{\rm HI}$ in the initial conditions, but the fact that these clouds maintain these column densities over time is promising. 

Intensity-weighted mean velocities (first spectral moment) and velocity dispersions (second moment) are more difficult to replicate. Mock observation velocity channels are initialized with their center at 0 km/s, and we shift the spectra by +80 km/s for comparison with observations. Therefore, when we compare $\overline{v}_{\rm LSR}$ of the Smith Cloud to mock observations, we focus on replicating the \textit{spread} of $\overline{v}_{\rm LSR}$. 

We were unable to fully match both the range of $\overline {v}_{\rm LSR}$ and the shape of the joint distributions involving it at our standard viewing angle of $\phi=90^\circ$. The mock observations, especially at later times, form more `blobby' distributions in the velocity and velocity dispersion parameter space (Figure~\ref{fig:moment1moment2}), with the center of the distribution around 80 km/s. Most mock observations also have velocity dispersions that are low compared to the Smith Cloud, though the overall distribution is similar (Figure~\ref{fig:moment0moment1}). The simulations at $\phi=90^\circ$ have a narrower velocity range that does not seem to correlate strongly with velocity dispersion.

From Figure~\ref{fig:m0m1_angles_3612}, decreasing the viewing angle $\phi$ to capture more of the cloud's bulk velocity shows promise in increasing the range of the line-of-sight velocity, similar to results from \citet{Henley2017}. In addition, not only do viewing angles of $\phi=30-60^\circ$ better match the observed range of $\overline{v}_{\rm LSR}$, but these are a better match to the predicted angle of $35-55^\circ$ \citep{Lockman2008}. These also reproduce the observed correlations with $\overline{v}_{\rm LSR}$, although only at early times before the cloud has been entrained, which is perhaps a hint as to the current dynamical state of the cloud.

Despite the lack of an exact match for the Smith Cloud, we elaborate on how patterns in these quantities may point to cloud evolution in Section~\ref{discussion_implications}. This is also likely influenced by several factors that we discuss further in Section~\ref{caveats}, including the omission of gravity and limitation on the observations' velocity channels. 

\subsubsection{Projected First-Order Velocity Structure Function}

Many of the projected first-order velocity structure functions of our simulations beyond $t=3t_{\rm cc}$ are similar to what is observed with GALFA-H\MakeUppercase{\romannumeral1} (Figures~\ref{fig:vsf} and \ref{fig:VSF_ACF_angles}).

Our closest match to the Smith Cloud is simulation C. It shows a good match at small separations, but, for our standard ($\phi =90^{\circ}$) viewing angle, the VSF flattens at large angles, which is not observed. However, the projected VSF at various viewing angles in Figure~\ref{fig:VSF_ACF_angles} shows that angles of $\phi=30-60^{\circ}$ do not exhibit this same flattening effect, instead becoming more similar to observations at later $t_{\rm cc}$, largely due to the inclusion of the cloud's bulk velocity along the line of sight. This distinction points to how both the small-scale turbulence within the cloud (low $\ell$) and the large-scale motions (large $\ell$) can be seen in the VCF.

Therefore, particularly when a cloud is growing and orientations are similar to what is quoted by \citet{Lockman2008}, a VSF quite similar to observations can be replicated as the cloud evolves. We discuss further considerations of the VSF in Section~\ref{discussion_VSF}.

\subsubsection{Normalized Autocovariance Function}

Even when varying simulation properties (metallicity, thermal pressure, cloud radius, and cooling) and mock observation parameters (viewing angles and cloud distance), none presented here are able to provide a perfect match for the normalized autocovariance function of neutral hydrogen column density, $\Gamma (N_{\rm HI})$, as calculated by Equation~\ref{acf_eqn}. In particular, while we can get good agreement on sub-degree scales, the simulations systematically underpredict the large-scale autocovariance.  

By comparing the observed and simulated autocovariance functions, we are able to hypothesize physical drivers that would better match the SC.   
First, recalling that $\Gamma(N_{\rm HI})\approx 1$ near low $\ell$ indicates nearly perfect correlation in $N_{\rm HI}$ between the closest pairs of points, we therefore expect that a larger cloud would exhibit less variation in column density, or (equivalently) have $\Gamma(N_{\rm HI})\approx 1$ for the same or larger $\ell$ than a smaller cloud. Similarly, if the function reaches a value of 0 around the cloud's radius, then this zero-point may indicate the cloud's size, indicating that a larger cloud can shift the ACF to larger scales.

It is not certain whether such clouds would have a similar slope to observations; we also note that the simulation initial conditions in this paper arise from an idealized scenario, such as a uniform cloud with constant velocity. It is possible that setting initial conditions from more physically motivated processes, such as generated by gravitational acceleration, may produce slopes similar to what is observed.
However, we note that this is only applicable for high-intensity clouds that are less sensitive to sigma clipping, such as for our simulation C. 

\subsection{Interpreting the Observed Velocity Structure Function}\label{discussion_VSF}

In this paper, we characterize various scales of turbulence in the Smith Cloud and in our mock observations through the projected first-order velocity structure function \citep[see also][]{Chira2019, Ha2021, Gronke2022, Hu2022, Chen2023, Abruzzo2024}. While the VSF can be useful for measuring turbulence on different scales, these works discuss considerations when using projected velocity structure functions, as most theoretical expectations are established in three dimensions. 

\citet{vonHoerner1951} demonstrated that the shape of the VSF may depend on the line-of-sight (LOS) cloud depth. Known as projection smoothing, this affects the slope of the VSF for separations smaller than the LOS cloud depth. For example, while idealized Kolmogorov turbulence is expected to have a slope of $\ell^{1/3}$, a steeper slope in the projected VSF might be recovered, especially at low $\ell$, with Kolmogorov turbulence being resolved at a higher $\ell$ \citep{vonHoerner1951, ODell1987, Xu2020, Mohapatra2022, Chen2023}. \citet{Li2020a} specifically argue that, in the context of galaxy clusters, the Kolmogorov slope is only recovered just before the turbulence driving scale. In this case, we do not expect to see Kolmogorov turbulence in these simulations due to underresolving the turbulent cascade \citep{Abruzzo2024, Rennehan2021, Mohapatra2022}. 

Furthermore, the VSF can indicate stages of cloud evolution. Because all of our simulated clouds begin as nearly uniform spheres with constant velocity, we might then expect VSFs to be similar at early times. Differences should appear as the cloud evolves, and they should correspond to the perceived cloud depth. 
Indeed, though our simulated clouds are slightly disrupted by the first cloud-crushing time, all early VSFs in Figure~\ref{fig:vsf} are similar. Changes in the slope are visible around the relative size of the cloud, which is larger for simulations A and B than C and D. Clouds A and B remain similar up to $5t_{\rm cc}$, where A's higher metallicity causes the evolution to differ and entrain more wind material. At later times for growing clouds A and C, their slopes change or become unstable at $\ell<R_{\rm cl}$, possibly indicating a change in the distribution of LOS cloud depths. 

As viewing angle --- and therefore cloud depth --- changes, so does the projection effect in Figure~\ref{fig:VSF_ACF_angles}. As $\phi$ varies, so does the $\ell$ of the slope transition and steepness. Low $\ell$ is unaffected, as this region probes turbulence within the cloud.  By $12t_{\rm cc}$ for all $\phi$, the slope is more constant due to the coevolution of the cloud and bulk velocity. Similar to results by \citet{Abruzzo2024} using the three-dimensional second-order VSF, we also find that the driving scale, or the peak of the VSF, changes very little over cloud-crushing times (for a given angle of $\phi)$.

Finally, we find it essential to address a significant result from \citet{Chen2023}: spatial smoothing directly impacts the VSF, and is more significant for flatter intrinsic slopes. Because the VSF is measured on spatial scales, spatial smoothing in RA/DEC may result in decreased $\langle| \delta v_\perp |\rangle$ at lower $\ell$, as velocity variations have been averaged over. This results in a steeper slope at low $\ell$, where the intrinsic VSF is not recovered until the highest spatial separations, if at all. While we spatially convolute our mock observations in order to match the observational beam size, we agree with \citet{Chen2023} in that any further spatial smoothing would result in better agreement at low $\ell$, but that this would be an artifact of little meaning as opposed to a true match. Similarly, this means that neglecting to match the mock observations' beam size to GALFA-HI would result in more disagreement of the VSF. For future studies, it is essential for comparison of the VSF that beam size is identical between observations and simulations, and if further spatial convolution is used (which we discourage), this should be accounted for by either noting the affected scales or implementing the correction of the second-order VSF by \citet{Chen2023}. 

\subsection{Implications for Observers} \label{discussion_implications}

We have shown a proof-of-concept demonstrating that, while not exactly matching the Smith Cloud, these comparison metrics may be used to investigate questions about observed clouds, including their evolutionary state, size, orientation, degree of turbulence, and their eventual fate. While it is difficult to use measurements at one point in time as a sole diagnostic for cloud survival, quantities in our mock observations appear to show several potentially useful patterns.

For example, growing clouds appear to have a slight drop in column density during entrainment, which later increases as the cloud accretes wind material. LOS velocity and velocity dispersion also show subtle variations depending on the cloud's stage of evolution. In simulation C, the ranges of $\overline{v}_{\rm LSR}$ and velocity dispersion increase as the cloud undergoes entrainment, then both decrease as the cloud continues growing. 

The first-order projected VSF also provides observers with potential information about turbulence driving at different scales, and through the projection effect (or lack thereof), it may be possible to constrain a cloud's orientation with respect to the bulk velocity and/or the evolution of this velocity along the length of the cloud. In the case of the Smith Cloud, this is more likely represented by an orientation that captures part of this motion within the line of sight, such as $\phi=30-60^{\circ}$ at late times (suggesting a more evolved cloud) in Figure~\ref{fig:VSF_ACF_angles}. 

An additional metric of cloud depth or radius (for high-intensity clouds) may also be the normalized autocovariance function, where values of $\Gamma(N_{\rm H\MakeUppercase{\romannumeral1}})\approx 1$ indicate an identical column density in the cloud at that particular separation $\ell$. As described in Section~\ref{acf}, the cloud's size significantly affects this, as well as where the function reaches zero. This makes intuitive sense: a larger radius results in a larger $\ell$ at which column densities are similar. For clouds that undergo steady growth, such as simulation C or both clouds in Appendix~\ref{resolution}, this function does not significantly evolve with cloud-crushing time, nor does it change with viewing angle, providing a more robust measurement. This metric may be more sensitive to HVC distance estimates, as the distance of a cloud affects the angular size $\ell$ on the sky. 

\subsection{Caveats and Considerations}\label{caveats}

As also noted in \citet{Abruzzo2024}, there are several considerations to take into account for both simulations and observations. 

From the simulation perspective, there are physical processes not implemented in the presented suite that may affect results such as cloud survival, radiative transfer, turbulence, and cloud morphologies. 
Perhaps the most significant missing physics is the implementation of gravity. At only $\sim$3 kpc from the Galactic plane \citep{Putman2003, Lockman2008}, the Smith Cloud's proximity means a strong gravitational force, and it has been observed to be falling into the Galaxy at an observed rate of $v_z=70$ km/s \citep{Lockman2008}. This provides an additional contribution to the cloud's overall velocity not simulated here, and as noted by \citet{Tan2023}, can have a significant impact on turbulence (and, by extension, the VSF), $\overline{v}_{\rm LSR}$, and velocity dispersions. Gravitational acceleration may drive the observed joint distributions that we fail to replicate (Figures~\ref{fig:moment0moment1} \& \ref{fig:moment1moment2}), as denser parts of the cloud will be less buoyant.

We also do not account for magnetic fields, which may defer (but not prevent) cloud destruction \citep{Dursi2008, McCourt2015}. Most relevant to this paper is their direct influence on turbulence, which \citet{Gronke2020} show may result in changes to cloud morphologies or lessening of hydrodynamical instabilities (e.g., Kelvin-Helmholtz) that propel mixing between the cloud and wind \citep{Galyardt2016, Abruzzo2024}. Self-shielding may also be relevant to reproducing observed column densities at lower pressures \citep{Sander2021, Farber2022}, and we acknowledge that the UV background from \citet{Haardt2012} does not take into account the SC's proximity to the Milky Way.  Finite resolution of the mock observations may also be something to consider: if a ray from radiative transfer encounters few cells ($R_{\rm cl}/\Delta x = 2, 4, 6,$ \textit{etc}), the resulting spectrum may be impacted.

Additional physics not included here are the possible presence of a dark matter halo \citep{Quilis2001, Nichols2009, Galyardt2016}, cosmic rays \citep{Butsky2020}, thermal conduction \citep{Sander2021}, and additional sources of turbulence, particularly in the simulation's initial conditions.

Finally, we note that recent X-ray observations show that the MW's virial and supervirial CGM phases may have significantly enhanced metallicity, resulting in solar and supersolar abundances for some species \citep{Das2019, Das2021, Gupta2021}. This could mean that HVCs such as the Smith Cloud are moving through a hot medium where TRML entrainment is even \textit{more} important, as increased metallicities result in more efficient cooling. 

On the observational side, the GALFA-H\MakeUppercase{\romannumeral1} data does not encompass the \textit{entire} Smith Cloud, reaching a minimum declination of $\delta=-0.725^{\circ}$, while the head of the cloud has been observed by \citet{Lockman2008} to reach $\delta \sim -2^{\circ}$ (Section~\ref{observationdetails}). Therefore, taking advantage of the much higher-resolution GALFA-H\MakeUppercase{\romannumeral1} data cubes results in the loss of part of the cloud. Upon imposing a similar limitation to the mock observations, we did not find any changes in our primary conclusions, which we find promising, though follow-ups with the lower-resolution data may prove beneficial. 

In terms of the observed $\overline{v}_{\rm LSR}$ range of the Smith Cloud, it appears that some SC emission may extend to velocity channels outside our imposed limit of $\sim75-130$ km/s. However, as some previous literature \citep{Stark2015, Minter2024}, this approximate cut is necessary to avoid Galactic contamination from the Galaxy's ISM that is difficult to filter out, but still captures the bulk of the SC's emission.

\section{Conclusions and Future Work}\label{conclusions}

In this paper, we present the first detailed comparison of an observed HVC (the Smith Cloud) to simulated clouds in the observational plane. To determine the simulations' ability to reproduce observations, we produced mock radio cubes and compared their column densities, average intensity-weighted LOS velocities, and intensity-weighted LOS velocity dispersions, characterized their degrees of turbulence through a projected first-order velocity structure function, and measured column densities across the clouds through a normalized autocovariance function. 

Although no mock observations are a match across all metrics for the Smith Cloud, we were able to reproduce many, and we examined physical drivers of these quantities and how they vary with metallicity (more efficient cooling), thermal pressure, initial size, cloud orientation, and distance.

Our primary conclusions can be summarized as follows:
\begin{enumerate}
    \item The chosen statical measures (zero, first, and second HI moment joint distributions, along with projected velocity structure functions and column density autocovariance functions) are effective in qualitatively comparing simulated and observed cloud properties. 

    \item While many of our simulations reproduce the range of observed moments and naturally reproduce the observed correlation between the HI column density and the velocity dispersion (Figure~\ref{fig:moment0moment2}), they generally do not reproduce the correlations between the other moments (Figure~\ref{fig:moment0moment1} and Figure~\ref{fig:moment1moment2}). The exception is simulation C with viewing angles $\phi=30-60^\circ$.

    \item Mock observations replicate the same range in average LOS velocity as seen in the Smith Cloud (Figure~\ref{fig:moment1moment2}) with viewing angles of $\phi=30-60^{\circ}$, similar to observational predictions for the SC's angle. Cloud orientation has a significant impact on the LOS velocity and projected VSF (Figures~\ref{fig:m0m1_angles_3612} and \ref{fig:VSF_ACF_angles})

    \item Given our assumed cloud initial conditions, the simulations imply that turbulent radiative mixing layers (TRML) and the physics of TRML entrainment are \textit{highly} relevant for HVCs, especially the Smith Cloud. This is key to match observed properties, such as the VSF (Figures~\ref{fig:vsf} and \ref{fig:VSF_ACF_angles}). This is evident in the evolution of clouds A and B, for example, which are identical aside from metallicity. 

    \item The projected VSF provides insight into cloud turbulence on small scales (low $\ell$) and bulk motion on larger scales (high $\ell$). The Smith Cloud maintains a nearly constant slope and extends to $\ell\sim10^\circ$, which simulation C only replicates at $\phi=30-60^\circ$. (Figures~\ref{fig:vsf} and \ref{fig:VSF_ACF_angles}.)
    
    \item No mock observations are a perfect match for the observed autocovariance function of column density on large scales. We suggest that the primary physical drivers of this function are the initial cloud size and whether the cloud is actively growing or being destroyed (see Figure~\ref{fig:2pc}).
    
\end{enumerate}

With this paper showing that these concepts and statistics may be applied (and even replicated) across radio data cubes of observed HVCs and mock observations, further work on this topic is critical to understanding the key physics involved in predicting the fate of the Milky Way's HVCs: whether they will fuel our Galaxy's hot halo medium or provide a source for future star formation. However, such future revelations will require significant contributions from both observations and simulations.

Hundreds of HVCs have been observed in the Milky Way, providing a large observational sample size through which to extend these comparisons. Follow-up work with different HVCs, particularly those with high-resolution observations, will be necessary to understand the robustness of these statistics beyond just the Smith Cloud. Future simulations should ideally include additional physics such as gravitational acceleration, and test various initial cloud setups, such as being initialized with a velocity gradient. These modifications may prove useful in better replicating the environments in which HVCs are formed and reside, and determine how the properties of clouds are affected.

\section*{Acknowledgements}

L.E.P. is grateful to David Schminovich, Frederik Paerels, and Kathryn Johnston for useful conversations about the manuscript. Authors ran simulations and analysis using Frontera allocation AST20007, supported by the NSF and TACC. The authors acknowledge support from NSF through grant AST-2307693. GLB acknowledges support from the NSF (AST-2108470, AST-2307419), NASA TCAN award 80NSSC21K1053, and the Simons Foundation through the Learning the Universe Collaboration. This research made use of \texttt{yt} \citep{Turk2011}, \texttt{matplotlib} \citep{Hunter2007}, \texttt{numpy} \citep{VanDerWalt2011}, and \texttt{scipy} \citep{Virtanen2020}.

\section*{Data Availability}

The data supporting the plots within this article are available on reasonable request to the corresponding author.



\bibliographystyle{mnras}
\bibliography{ref} 

\begin{thebibliography}{}
\makeatletter
\relax
\def\mn@urlcharsother{\let\do\@makeother \do\$\do\&\do\#\do\^\do\_\do\%\do\~}
\def\mn@doi{\begingroup\mn@urlcharsother \@ifnextchar [ {\mn@doi@} {\mn@doi@[]}}
\def\mn@doi@[#1]#2{\def\@tempa{#1}\ifx\@tempa\@empty \href {http://dx.doi.org/#2} {doi:#2}\else \href {http://dx.doi.org/#2} {#1}\fi \endgroup}
\def\mn@eprint#1#2{\mn@eprint@#1:#2::\@nil}
\def\mn@eprint@arXiv#1{\href {http://arxiv.org/abs/#1} {{\tt arXiv:#1}}}
\def\mn@eprint@dblp#1{\href {http://dblp.uni-trier.de/rec/bibtex/#1.xml} {dblp:#1}}
\def\mn@eprint@#1:#2:#3:#4\@nil{\def\@tempa {#1}\def\@tempb {#2}\def\@tempc {#3}\ifx \@tempc \@empty \let \@tempc \@tempb \let \@tempb \@tempa \fi \ifx \@tempb \@empty \def\@tempb {arXiv}\fi \@ifundefined {mn@eprint@\@tempb}{\@tempb:\@tempc}{\expandafter \expandafter \csname mn@eprint@\@tempb\endcsname \expandafter{\@tempc}}}

\bibitem[\protect\citeauthoryear{Abruzzo, Bryan  \& Fielding}{Abruzzo et~al.}{2022}]{Abruzzo2022}
Abruzzo M.~W.,  Bryan G.~L.,   Fielding D.~B.,  2022, \mn@doi [Astrophys. J.] {10.3847/1538-4357/ac3c48}, 925, 199

\bibitem[\protect\citeauthoryear{Abruzzo, Fielding  \& Bryan}{Abruzzo et~al.}{2023}]{Abruzzo2023}
Abruzzo M.~W.,  Fielding D.~B.,   Bryan G.~L.,  2023, \mn@doi [ArXiv eprints] {https://ui.adsabs.harvard.edu/link_gateway/2023arXiv230703228A/doi:10.48550/arXiv.2307.03228}

\bibitem[\protect\citeauthoryear{Abruzzo, Fielding  \& Bryan}{Abruzzo et~al.}{2024}]{Abruzzo2024}
Abruzzo M.~W.,  Fielding D.~B.,   Bryan G.~L.,  2024, \mn@doi [Astrophys. J.] {10.3847/1538-4357/ad1e51}, 966, 181

\bibitem[\protect\citeauthoryear{Armillotta, Fraternali  \& Marinacci}{Armillotta et~al.}{2016}]{Armillotta2016}
Armillotta L.,  Fraternali F.,   Marinacci F.,  2016, \mn@doi [Mon. Not. R. Astron. Soc.] {10.1093/mnras/stw1930}, 462, 4157

\bibitem[\protect\citeauthoryear{Balbus \& McKee}{Balbus \& McKee}{1982}]{Balbus1982}
Balbus S.,  McKee C.~F.,  1982, \mn@doi [Astrophys. J.] {https://ui.adsabs.harvard.edu/link_gateway/1982ApJ...252..529B/doi:10.1086/159581}, 252, 529

\bibitem[\protect\citeauthoryear{{Ben Bekhti}, Richter, Winkel, Kenn  \& Westmeier}{{Ben Bekhti} et~al.}{2009}]{BenBekhti2009}
{Ben Bekhti} N.,  Richter P.,  Winkel B.,  Kenn F.,   Westmeier T.,  2009, \mn@doi [Astron. Astrophys.] {10.1051/0004-6361/200811259}, 503, 483

\bibitem[\protect\citeauthoryear{Bordner \& Norman}{Bordner \& Norman}{2012}]{BordnerNorman2012}
Bordner J.,  Norman M.~L.,  2012, in BW-XSEDE'12 Proc. Extrem. Scaling Work.. Champaign, IL, pp 1--11

\bibitem[\protect\citeauthoryear{Bordner \& Norman}{Bordner \& Norman}{2018}]{BordnerNorman2018}
Bordner J.,  Norman M.~L.,  2018, \mn@doi [ArXiv eprints] {https://ui.adsabs.harvard.edu/link_gateway/2018arXiv181001319B/doi:10.48550/arXiv.1810.01319}

\bibitem[\protect\citeauthoryear{Br{\"{u}}ns, Kerp, Kalberla  \& Mebold}{Br{\"{u}}ns et~al.}{2000}]{Bruns2000}
Br{\"{u}}ns C.,  Kerp J.,  Kalberla P.~M.,   Mebold U.,  2000, \mn@doi [Astron. Astrophys.] {https://ui.adsabs.harvard.edu/link_gateway/2000A&A...357..120B/doi:10.48550/arXiv.astro-ph/0003110}, 357, 120

\bibitem[\protect\citeauthoryear{Bryan et~al.,}{Bryan et~al.}{2014}]{Bryan2014}
Bryan G.~L.,  et~al., 2014, \mn@doi [Astrophys. Journal, Suppl. Ser.] {10.1088/0067-0049/211/2/19}, 211

\bibitem[\protect\citeauthoryear{Bustard \& Gronke}{Bustard \& Gronke}{2022}]{Bustard2022}
Bustard C.,  Gronke M.,  2022, \mn@doi [Astrophys. J.] {10.3847/1538-4357/ac752b}, 933, 120

\bibitem[\protect\citeauthoryear{Butsky, Fielding, Hayward, Hummels, Quinn  \& Werk}{Butsky et~al.}{2020}]{Butsky2020}
Butsky I.~S.,  Fielding D.~B.,  Hayward C.~C.,  Hummels C.~B.,  Quinn T.~R.,   Werk J.~K.,  2020, \mn@doi [Astrophys. J.] {10.3847/1538-4357/abbad2}, 903, 77

\bibitem[\protect\citeauthoryear{Chen \& {Peng Oh}}{Chen \& {Peng Oh}}{2024}]{Chen2024}
Chen Z.,  {Peng Oh} S.,  2024, \mn@doi [Mon. Not. R. Astron. Soc.] {https://doi.org/10.1093/mnras/stae1113}, 530, 4032

\bibitem[\protect\citeauthoryear{Chen et~al.,}{Chen et~al.}{2023a}]{Chen2023}
Chen M.~C.,  et~al., 2023a, \mn@doi [Mon. Not. R. Astron. Soc.] {10.1093/mnras/stac3193}, 518, 2354

\bibitem[\protect\citeauthoryear{Chen, Fielding  \& Bryan}{Chen et~al.}{2023b}]{Chen2023b}
Chen Z.,  Fielding D.~B.,   Bryan G.~L.,  2023b, \mn@doi [Astrophys. J.] {10.3847/1538-4357/acc73f}, 950, 91

\bibitem[\protect\citeauthoryear{Chiappini, Matteucci  \& Romano}{Chiappini et~al.}{2001}]{Chiappini2001}
Chiappini C.,  Matteucci F.,   Romano D.,  2001, \mn@doi [Astrophys. J.] {10.1086/321427}, 554, 1044

\bibitem[\protect\citeauthoryear{Chiappini, Matteucci  \& Meynet}{Chiappini et~al.}{2003}]{Chiappini2003}
Chiappini C.,  Matteucci F.,   Meynet G.,  2003, \mn@doi [Astron. Astrophys.] {10.1051/0004-6361:20031192}, 410, 257

\bibitem[\protect\citeauthoryear{Chira, Ib{\'{a}}{\~{n}}ez-Mej{\'{i}}a, {Mac Low}  \& Henning}{Chira et~al.}{2019}]{Chira2019}
Chira R.~A.,  Ib{\'{a}}{\~{n}}ez-Mej{\'{i}}a J.~C.,  {Mac Low} M.~M.,   Henning T.,  2019, \mn@doi [Astron. Astrophys.] {10.1051/0004-6361/201833970}, 630, 1

\bibitem[\protect\citeauthoryear{Chomiuk \& Povich}{Chomiuk \& Povich}{2011}]{Chomiuk2011}
Chomiuk L.,  Povich M.~S.,  2011, \mn@doi [Astron. J.] {10.1088/0004-6256/142/6/197}, 142

\bibitem[\protect\citeauthoryear{Cooper, Bicknell, Sutherland  \& Bland-Hawthorn}{Cooper et~al.}{2009}]{Cooper2009}
Cooper J.~L.,  Bicknell G.~V.,  Sutherland R.~S.,   Bland-Hawthorn J.,  2009, \mn@doi [Astrophys. J.] {10.1088/0004-637X/703/1/330}, 703, 330

\bibitem[\protect\citeauthoryear{Das, Mathur, Nicastro  \& Krongold}{Das et~al.}{2019}]{Das2019}
Das S.,  Mathur S.,  Nicastro F.,   Krongold Y.,  2019, \mn@doi [Astrophys. J. Lett.] {10.3847/2041-8213/ab3b09}, 882, L23

\bibitem[\protect\citeauthoryear{Das, Mathur, Gupta  \& Krongold}{Das et~al.}{2021}]{Das2021}
Das S.,  Mathur S.,  Gupta A.,   Krongold Y.,  2021, \mn@doi [Astrophys. J.] {10.3847/1538-4357/ac0e8e}, 918, 83

\bibitem[\protect\citeauthoryear{Dav{\'{e}}, Angl{\'{e}}s-Alc{\'{a}}zar, Narayanan, Li, Rafieferantsoa  \& Appleby}{Dav{\'{e}} et~al.}{2019}]{Dave2019}
Dav{\'{e}} R.,  Angl{\'{e}}s-Alc{\'{a}}zar D.,  Narayanan D.,  Li Q.,  Rafieferantsoa M.~H.,   Appleby S.,  2019, \mn@doi [Mon. Not. R. Astron. Soc.] {10.1093/mnras/stz937}, 486, 2827

\bibitem[\protect\citeauthoryear{Dursi \& Pfrommer}{Dursi \& Pfrommer}{2008}]{Dursi2008}
Dursi L.~J.,  Pfrommer C.,  2008, \mn@doi [Astrophys. J.] {10.1086/529371}, 677, 993

\bibitem[\protect\citeauthoryear{Elia et~al.,}{Elia et~al.}{2022}]{Elia2022}
Elia D.,  et~al., 2022, \mn@doi [Astrophys. J.] {10.3847/1538-4357/aca27d}, 941, 162

\bibitem[\protect\citeauthoryear{Erb}{Erb}{2008}]{Erb2008}
Erb D.~K.,  2008, \mn@doi [Astrophys. J.] {10.1086/524727}, 674, 151

\bibitem[\protect\citeauthoryear{Farber, Gronke, Planck  \& M}{Farber et~al.}{2022}]{Farber2022}
Farber R.~J.,  Gronke M.,  Planck M.,   M D.-G.,  2022, \mn@doi [Mon. Not. R. Astron. Soc.] {https://ui.adsabs.harvard.edu/link_gateway/2022MNRAS.510..551F/doi:10.1093/mnras/stab3412}, 510, 551

\bibitem[\protect\citeauthoryear{Federrath}{Federrath}{2013}]{Federrath2013}
Federrath C.,  2013, \mn@doi [Mon. Not. R. Astron. Soc.] {10.1093/mnras/stt1644}, 436, 1245

\bibitem[\protect\citeauthoryear{Fielding, Ostriker, Bryan  \& Jermyn}{Fielding et~al.}{2020}]{Fielding2020}
Fielding D.~B.,  Ostriker E.~C.,  Bryan G.~L.,   Jermyn A.~S.,  2020, \mn@doi [Astrophys. J. Lett.] {10.3847/2041-8213/ab8d2c}, 894, L24

\bibitem[\protect\citeauthoryear{Fox et~al.,}{Fox et~al.}{2014}]{Fox2014}
Fox A.~J.,  et~al., 2014, \mn@doi [Astrophys. J.] {10.1088/0004-637X/787/2/147}, 787

\bibitem[\protect\citeauthoryear{Fox et~al.,}{Fox et~al.}{2016}]{Fox2016}
Fox A.~J.,  et~al., 2016, \mn@doi [Astrophys. J. Lett.] {10.3847/2041-8205/816/1/l11}, 816, L11

\bibitem[\protect\citeauthoryear{Fox, Richter, Ashley, Heckman, Lehner, Werk, Bordoloi  \& Peeples}{Fox et~al.}{2019}]{Fox2019}
Fox A.~J.,  Richter P.,  Ashley T.,  Heckman T.~M.,  Lehner N.,  Werk J.~K.,  Bordoloi R.,   Peeples M.~S.,  2019, \mn@doi [Astrophys. J.] {10.3847/1538-4357/ab40ad}, 884, 53

\bibitem[\protect\citeauthoryear{Frisch}{Frisch}{1995}]{Frisch1995}
Frisch U.,  1995, {Turbulence: The legacy of A.N. Kolmogorov}.
Cambridge University Press., Cambridge

\bibitem[\protect\citeauthoryear{Fuchs, Jahrei  \& Flynn}{Fuchs et~al.}{2009}]{Fuchs2009}
Fuchs B.,  Jahrei H.,   Flynn C.,  2009, \mn@doi [Astron. J.] {10.1088/0004-6256/137/1/266}, 137, 266

\bibitem[\protect\citeauthoryear{Galyardt \& Shelton}{Galyardt \& Shelton}{2016}]{Galyardt2016}
Galyardt J.,  Shelton R.~L.,  2016, \mn@doi [Astrophys. J. Lett.] {10.3847/2041-8205/816/1/l18}, 816, L18

\bibitem[\protect\citeauthoryear{Gronke \& Oh}{Gronke \& Oh}{2018}]{Gronke2018}
Gronke M.,  Oh S.~P.,  2018, \mn@doi [Mon. Not. R. Astron. Soc. Lett.] {10.1093/mnrasl/sly131}, 480, L111

\bibitem[\protect\citeauthoryear{Gronke \& Oh}{Gronke \& Oh}{2020a}]{Gronke2020a}
Gronke M.,  Oh S.~P.,  2020a, \mn@doi [Mon. Not. R. Astron. Soc.] {10.1093/mnras/stz3332}, 492, 1970

\bibitem[\protect\citeauthoryear{Gronke \& Oh}{Gronke \& Oh}{2020b}]{Gronke2020}
Gronke M.,  Oh S.~P.,  2020b, \mn@doi [Mon. Not. R. Astron. Soc. Lett.] {10.1093/mnrasl/slaa033}, 494, L27

\bibitem[\protect\citeauthoryear{Gronke, Oh, Ji  \& Norman}{Gronke et~al.}{2022}]{Gronke2022}
Gronke M.,  Oh S.~P.,  Ji S.,   Norman C.,  2022, \mn@doi [Mon. Not. R. Astron. Soc.] {https://ui.adsabs.harvard.edu/link_gateway/2022MNRAS.511..859G/doi:10.1093/mnras/stab3351}, 511, 859

\bibitem[\protect\citeauthoryear{Gupta, Kingsbury, Mathur, Das, Galeazzi, Krongold  \& Nicastro}{Gupta et~al.}{2021}]{Gupta2021}
Gupta A.,  Kingsbury J.,  Mathur S.,  Das S.,  Galeazzi M.,  Krongold Y.,   Nicastro F.,  2021, \mn@doi [Astrophys. J.] {10.3847/1538-4357/abdbb6}, 909, 164

\bibitem[\protect\citeauthoryear{Ha, Li, Xu, Kounkel  \& Li}{Ha et~al.}{2021}]{Ha2021}
Ha T.,  Li Y.,  Xu S.,  Kounkel M.,   Li H.,  2021, \mn@doi [Astrophys. J. Lett.] {10.3847/2041-8213/abd8c9}, 907, L40

\bibitem[\protect\citeauthoryear{Haardt \& Madau}{Haardt \& Madau}{2012}]{Haardt2012}
Haardt F.,  Madau P.,  2012, \mn@doi [Astrophys. J.] {10.1088/0004-637X/746/2/125}, 746

\bibitem[\protect\citeauthoryear{Heitsch \& Putman}{Heitsch \& Putman}{2009}]{Heitsch2009}
Heitsch F.,  Putman M.~E.,  2009, \mn@doi [Astrophys. J.] {10.1088/0004-637X/698/2/1485}, 698, 1485

\bibitem[\protect\citeauthoryear{Heitsch, Bartell, Clark, Peek, Cheng  \& Putman}{Heitsch et~al.}{2016}]{Heitsch2016}
Heitsch F.,  Bartell B.,  Clark S.~E.,  Peek J.~E.,  Cheng D.,   Putman M.,  2016, \mn@doi [Mon. Not. R. Astron. Soc. Lett.] {10.1093/mnrasl/slw124}, 462, L46

\bibitem[\protect\citeauthoryear{Heitsch, Marchal, Miville-Desch{\^{e}}nes, Shull  \& Fox}{Heitsch et~al.}{2022}]{Heitsch2022}
Heitsch F.,  Marchal A.,  Miville-Desch{\^{e}}nes M.~A.,  Shull J.~M.,   Fox A.~J.,  2022, \mn@doi [Mon. Not. R. Astron. Soc.] {10.1093/mnras/stab3266}, 509, 4515

\bibitem[\protect\citeauthoryear{Henley, Gritton  \& Shelton}{Henley et~al.}{2017}]{Henley2017}
Henley D.~B.,  Gritton J.~A.,   Shelton R.~L.,  2017, \mn@doi [Astrophys. J.] {10.3847/1538-4357/aa5df7}, 837, 82

\bibitem[\protect\citeauthoryear{Hidalgo-Pineda, Farber  \& Gronke}{Hidalgo-Pineda et~al.}{2024}]{Hidalgo-Pineda2024}
Hidalgo-Pineda F.,  Farber R.~J.,   Gronke M.,  2024, \mn@doi [Mon. Not. R. Astron. Soc.] {https://doi.org/10.3847/1538-4357/acc73f}, 527, 135

\bibitem[\protect\citeauthoryear{Hill, Haffner  \& Reynolds}{Hill et~al.}{2009}]{Hill2009}
Hill A.~S.,  Haffner L.~M.,   Reynolds R.~J.,  2009, \mn@doi [Astrophys. J.] {10.1088/0004-637X/703/2/1832}, 703, 1832

\bibitem[\protect\citeauthoryear{Holm-hansen, Putman  \& Kim}{Holm-hansen et~al.}{2025}]{Holmhansen2025}
Holm-hansen C.,  Putman M.~E.,   Kim D.~A.,  2025, \mn@doi [Mon. Not. R. Astron. Soc.] {https://ui.adsabs.harvard.edu/link_gateway/2025MNRAS.536.3507H/doi:10.1093/mnras/stae2801}, 536, 3507

\bibitem[\protect\citeauthoryear{Hopkins, McClure-Griffiths  \& Gaensler}{Hopkins et~al.}{2008}]{Hopkins2008}
Hopkins A.~M.,  McClure-Griffiths N.~M.,   Gaensler B.~M.,  2008, \mn@doi [Astrophys. J.] {10.1086/590494}, 682, L13

\bibitem[\protect\citeauthoryear{Hsu, Putman, Heitsch, Stanimirovi{\'{c}}, Peek  \& Clark}{Hsu et~al.}{2011}]{Hsu2011}
Hsu W.~H.,  Putman M.~E.,  Heitsch F.,  Stanimirovi{\'{c}} S.,  Peek J.~E.,   Clark S.~E.,  2011, \mn@doi [Astron. J.] {10.1088/0004-6256/141/2/57}, 141

\bibitem[\protect\citeauthoryear{Hu, Federrath, Xu  \& Mathew}{Hu et~al.}{2022}]{Hu2022}
Hu Y.,  Federrath C.,  Xu S.,   Mathew S.~S.,  2022, \mn@doi [Mon. Not. R. Astron. Soc.] {10.1093/mnras/stac972}, 513, 2100

\bibitem[\protect\citeauthoryear{Hunter}{Hunter}{2007}]{Hunter2007}
Hunter J.~D.,  2007, \mn@doi [Comput. Sci. Eng.] {https://doi.org/10.1109/MCSE.2007.55}, 9, 90

\bibitem[\protect\citeauthoryear{Ji, Oh  \& Masterson}{Ji et~al.}{2019}]{Ji2019}
Ji S.,  Oh S.~P.,   Masterson P.,  2019, \mn@doi [Mon. Not. R. Astron. Soc.] {10.1093/mnras/stz1248}, 487, 737

\bibitem[\protect\citeauthoryear{Joung, Bryan  \& Putman}{Joung et~al.}{2012}]{Joung2012}
Joung M.~R.,  Bryan G.~L.,   Putman M.~E.,  2012, \mn@doi [Astrophys. J.] {10.1088/0004-637X/745/2/148}, 745

\bibitem[\protect\citeauthoryear{Kanjilal, Dutta  \& Sharma}{Kanjilal et~al.}{2021}]{Kanjilal2021}
Kanjilal V.,  Dutta A.,   Sharma P.,  2021, \mn@doi [Mon. Not. R. Astron. Soc.] {10.1093/mnras/staa3610}, 501, 1143

\bibitem[\protect\citeauthoryear{Kennicutt \& Evans}{Kennicutt \& Evans}{2012}]{Kennicutt2012}
Kennicutt R.~C.,  Evans N.~J.,  2012, \mn@doi [Annu. Rev. Astron. Astrophys.] {10.1146/annurev-astro-081811-125610}, 50, 531

\bibitem[\protect\citeauthoryear{Klein, McKee  \& Colella}{Klein et~al.}{1994}]{Klein1994}
Klein R.,  McKee C.~F.,   Colella P.,  1994, \mn@doi [Astrophys. J.] {https://ui.adsabs.harvard.edu/link_gateway/1994ApJ...420..213K/doi:10.1086/173554}, 420, 213

\bibitem[\protect\citeauthoryear{Larson, Tinsley  \& Caldwell}{Larson et~al.}{1980}]{Larson1980}
Larson R.~B.,  Tinsley B.,   Caldwell N.,  1980, \mn@doi [Astrophys. J.] {https://ui.adsabs.harvard.edu/link_gateway/1980ApJ...237..692L/doi:10.1086/157917}, 237, 692

\bibitem[\protect\citeauthoryear{Lehner, Howk, Marasco  \& Fraternali}{Lehner et~al.}{2022}]{Lehner2022}
Lehner N.,  Howk J.~C.,  Marasco A.,   Fraternali F.,  2022, \mn@doi [Mon. Not. R. Astron. Soc.] {https://doi.org/10.1093/mnras/stac987}, 513, 3228

\bibitem[\protect\citeauthoryear{Li, Hopkins, Squire  \& Hummels}{Li et~al.}{2020a}]{Li2020b}
Li Z.,  Hopkins P.~F.,  Squire J.,   Hummels C.,  2020a, \mn@doi [Mon. Not. R. Astron. Soc.] {10.1093/mnras/stz3567}, 492, 1841

\bibitem[\protect\citeauthoryear{Li et~al.,}{Li et~al.}{2020b}]{Li2020a}
Li Y.,  et~al., 2020b, \mn@doi [Astrophys. J. Lett.] {10.3847/2041-8213/ab65c7}, 889, L1

\bibitem[\protect\citeauthoryear{Li, Luo, Fossati, Sun, J  \& Occhialini}{Li et~al.}{2023}]{Li2023}
Li Y.,  Luo R.,  Fossati M.,  Sun M.,  J P.,   Occhialini F.~G.,  2023, \mn@doi [Mon. Not. R. Astron. Soc.] {https://doi.org/10.1093/mnras/stad874}, 521, 4785

\bibitem[\protect\citeauthoryear{Licquia \& Newman}{Licquia \& Newman}{2015}]{Licquia2015}
Licquia T.~C.,  Newman J.~A.,  2015, \mn@doi [Astrophys. J.] {10.1088/0004-637X/806/1/96}, 806, 96

\bibitem[\protect\citeauthoryear{Lockman, Benjamin, Heroux  \& Langston}{Lockman et~al.}{2008}]{Lockman2008}
Lockman F.~J.,  Benjamin R.~A.,  Heroux A.~J.,   Langston G.~I.,  2008, \mn@doi [Astrophys. J.] {10.1086/588838}, 679, L21

\bibitem[\protect\citeauthoryear{Maller \& Bullock}{Maller \& Bullock}{2004}]{Maller2004}
Maller A.~H.,  Bullock J.~S.,  2004, \mn@doi [Mon. Not. R. Astron. Soc.] {10.1111/j.1365-2966.2004.08349.x}, 355, 694

\bibitem[\protect\citeauthoryear{Marinacci, Binney, Fraternali, Nipoti, Ciotti  \& Londrillo}{Marinacci et~al.}{2010}]{Marinacci2010}
Marinacci F.,  Binney J.,  Fraternali F.,  Nipoti C.,  Ciotti L.,   Londrillo P.,  2010, \mn@doi [Mon. Not. R. Astron. Soc.] {10.1111/j.1365-2966.2010.16352.x}, 404, 1464

\bibitem[\protect\citeauthoryear{McCourt, O'Leary, Madigan  \& Quataert}{McCourt et~al.}{2015}]{McCourt2015}
McCourt M.,  O'Leary R.~M.,  Madigan A.~M.,   Quataert E.,  2015, \mn@doi [Mon. Not. R. Astron. Soc.] {10.1093/mnras/stv355}, 449, 2

\bibitem[\protect\citeauthoryear{McKee \& Cowie}{McKee \& Cowie}{1977}]{McKee1977}
McKee C.~F.,  Cowie L.~L.,  1977, \mn@doi [Astrophys. J.] {https://ui.adsabs.harvard.edu/link_gateway/1977ApJ...215..213M/doi:10.1086/155350}, 215, 213

\bibitem[\protect\citeauthoryear{Minter, Lockman, Balashev  \& Ford}{Minter et~al.}{2024}]{Minter2024}
Minter A.~H.,  Lockman F.~J.,  Balashev S.~A.,   Ford H.~A.,  2024, \mn@doi [Astrophys. J.] {10.3847/1538-4357/ad343d}, 966, 76

\bibitem[\protect\citeauthoryear{Mohapatra, Jetti, Sharma  \& Federrath}{Mohapatra et~al.}{2022}]{Mohapatra2022}
Mohapatra R.,  Jetti M.,  Sharma P.,   Federrath C.,  2022, \mn@doi [Mon. Not. R. Astron. Soc.] {10.1093/mnras/stab3429}, 510, 2327

\bibitem[\protect\citeauthoryear{Nichols \& Bland-Hawthorn}{Nichols \& Bland-Hawthorn}{2009}]{Nichols2009}
Nichols M.,  Bland-Hawthorn J.,  2009, \mn@doi [Astrophys. J.] {10.1088/0004-637X/707/2/1642}, 707, 1642

\bibitem[\protect\citeauthoryear{O'Dell \& Castaneda}{O'Dell \& Castaneda}{1987}]{ODell1987}
O'Dell C.~R.,  Castaneda H.~O.,  1987, \mn@doi [Astrophys. J.] {https://ui.adsabs.harvard.edu/link_gateway/1987ApJ...317..686O/doi:10.1086/165314}, 317, 686

\bibitem[\protect\citeauthoryear{Oort}{Oort}{1969}]{Oort1969}
Oort J.~H.,  1969, \mn@doi [Nature] {10.1038/2241158a0}, 224, 1158

\bibitem[\protect\citeauthoryear{Ossenkopf \& Low}{Ossenkopf \& Low}{2002}]{Ossenkopf2002}
Ossenkopf V.,  Low M.~M.,  2002, \mn@doi [Astron. Astrophys.] {10.1051/0004-6361:20020629}, 390, 307

\bibitem[\protect\citeauthoryear{Peek, Putman, McKee, Heiles  \& Stanimirovi{\'{c}}}{Peek et~al.}{2007}]{Peek2007}
Peek J. E.~G.,  Putman M.~E.,  McKee C.~F.,  Heiles C.,   Stanimirovi{\'{c}} S.,  2007, \mn@doi [Astrophys. J.] {10.1086/510189}, 656, 907

\bibitem[\protect\citeauthoryear{Peek et~al.,}{Peek et~al.}{2018}]{Peek2018}
Peek J. E.~G.,  et~al., 2018, \mn@doi [Astrophys. J. Suppl. Ser.] {10.3847/1538-4365/aa91d3}, 234, 2

\bibitem[\protect\citeauthoryear{Pillepich et~al.,}{Pillepich et~al.}{2018}]{Pillepich2018}
Pillepich A.,  et~al., 2018, \mn@doi [Mon. Not. R. Astron. Soc.] {10.1093/mnras/stx2656}, 473, 4077

\bibitem[\protect\citeauthoryear{Putman, Bland-Hawthorn, Veilleux, Gibson, Freeman  \& Maloney}{Putman et~al.}{2003}]{Putman2003}
Putman M.~E.,  Bland-Hawthorn J.,  Veilleux S.,  Gibson B.~K.,  Freeman K.~C.,   Maloney P.~R.,  2003, \mn@doi [Astrophys. J.] {https://ui.adsabs.harvard.edu/link_gateway/2003ApJ...597..948P/doi:10.1086/378555}, 597, 948

\bibitem[\protect\citeauthoryear{Putman, Peek  \& Joung}{Putman et~al.}{2012}]{Putman2012}
Putman M.~E.,  Peek J.~E.,   Joung M.~R.,  2012, \mn@doi [Annu. Rev. Astron. Astrophys.] {10.1146/annurev-astro-081811-125612}, 50, 491

\bibitem[\protect\citeauthoryear{Quilis \& Moore}{Quilis \& Moore}{2001}]{Quilis2001}
Quilis V.,  Moore B.,  2001, \mn@doi [Astrophys. J.] {10.1086/322866}, 555, L95

\bibitem[\protect\citeauthoryear{Rennehan}{Rennehan}{2021}]{Rennehan2021}
Rennehan D.,  2021, \mn@doi [Mon. Not. R. Astron. Soc.] {10.1093/mnras/stab1813}, 506, 2836

\bibitem[\protect\citeauthoryear{Richie, Schneider, Abruzzo  \& Torrey}{Richie et~al.}{2024}]{Richie2024}
Richie H.~M.,  Schneider E.~E.,  Abruzzo M.~W.,   Torrey P.,  2024, \mn@doi [Astrophys. J.] {https://ui.adsabs.harvard.edu/link_gateway/2024ApJ...974...81R/doi:10.3847/1538-4357/ad6a1c}, 974, 81

\bibitem[\protect\citeauthoryear{Robitaille \& Whitney}{Robitaille \& Whitney}{2010}]{Robitaille2010}
Robitaille T.~P.,  Whitney B.~A.,  2010, \mn@doi [Astrophys. J. Lett.] {10.1088/2041-8205/710/1/L11}, 710, 11

\bibitem[\protect\citeauthoryear{Sander \& Hensler}{Sander \& Hensler}{2021}]{Sander2021}
Sander B.,  Hensler G.,  2021, \mn@doi [Mon. Not. R. Astron. Soc.] {10.1093/mnras/staa3952}, 501, 5330

\bibitem[\protect\citeauthoryear{Scannapieco \& Br{\"{u}}ggen}{Scannapieco \& Br{\"{u}}ggen}{2015}]{Scannapieco2015}
Scannapieco E.,  Br{\"{u}}ggen M.,  2015, \mn@doi [Astrophys. J.] {10.1088/0004-637X/805/2/158}, 805, 158

\bibitem[\protect\citeauthoryear{Schneider \& Robertson}{Schneider \& Robertson}{2017}]{Schneider2017}
Schneider E.~E.,  Robertson B.~E.,  2017, \mn@doi [Astrophys. J.] {10.3847/1538-4357/834/2/144}, 834, 144

\bibitem[\protect\citeauthoryear{Sembach et~al.,}{Sembach et~al.}{2003}]{Sembach2003}
Sembach K.~R.,  et~al., 2003, \mn@doi [Astrophys. J. Suppl. Ser.] {10.1086/346231}, 146, 165

\bibitem[\protect\citeauthoryear{Smith}{Smith}{1963}]{Smith1963}
Smith G.,  1963, \mn@doi [Bull. Astron. Institutes Netherlands] {https://ui.adsabs.harvard.edu/abs/1963BAN....17..203S/abstract}, 17, 203

\bibitem[\protect\citeauthoryear{Smith et~al.,}{Smith et~al.}{2017}]{Smith2017}
Smith B.~D.,  et~al., 2017, \mn@doi [Mon. Not. R. Astron. Soc.] {10.1093/mnras/stw3291}, 466, 2217

\bibitem[\protect\citeauthoryear{Sparre, Pfrommer  \& Vogelsberger}{Sparre et~al.}{2019}]{Sparre2019}
Sparre M.,  Pfrommer C.,   Vogelsberger M.,  2019, \mn@doi [Mon. Not. R. Astron. Soc.] {10.1093/mnras/sty3063}, 482, 5401

\bibitem[\protect\citeauthoryear{Sparre, Pfrommer  \& Ehlert}{Sparre et~al.}{2020}]{Sparre2020}
Sparre M.,  Pfrommer C.,   Ehlert K.,  2020, \mn@doi [Mon. Not. R. Astron. Soc.] {10.1093/mnras/staa3177}, 499, 4261

\bibitem[\protect\citeauthoryear{Stanimirovi{\'{c}} et~al.,}{Stanimirovi{\'{c}} et~al.}{2006}]{Stanimirovic2006}
Stanimirovi{\'{c}} S.,  et~al., 2006, \mn@doi [Astrophys. J.] {10.1086/508800}, 653, 1210

\bibitem[\protect\citeauthoryear{Stark, Baker  \& Kannappan}{Stark et~al.}{2015}]{Stark2015}
Stark D.~V.,  Baker A.~D.,   Kannappan S.~J.,  2015, \mn@doi [Mon. Not. R. Astron. Soc.] {10.1093/mnras/stu2182}, 446, 1855

\bibitem[\protect\citeauthoryear{Tan, {Peng Oh}  \& Gronke}{Tan et~al.}{2021}]{Tan2021}
Tan B.,  {Peng Oh} S.,   Gronke M.,  2021, \mn@doi [Mon. Not. R. Astron. Soc.] {10.1093/mnras/stab053}, 502, 3179

\bibitem[\protect\citeauthoryear{Tan, Oh  \& Gronke}{Tan et~al.}{2023}]{Tan2023}
Tan B.,  Oh S.~P.,   Gronke M.,  2023, \mn@doi [Mon. Not. R. Astron. Soc.] {https://ui.adsabs.harvard.edu/link_gateway/2023MNRAS.520.2571T/doi:10.1093/mnras/stad236}, 520, 2571

\bibitem[\protect\citeauthoryear{Turk, Smith, Oishi, Skory, Skillman, Abel  \& Norman}{Turk et~al.}{2011}]{Turk2011}
Turk M.~J.,  Smith B.~D.,  Oishi J.~S.,  Skory S.,  Skillman S.~W.,  Abel T.,   Norman M.~L.,  2011, \mn@doi [Astrophys. Journal, Suppl. Ser.] {10.1088/0067-0049/192/1/9}, 192

\bibitem[\protect\citeauthoryear{{Van Der Walt}, Colbert  \& Varoquaux}{{Van Der Walt} et~al.}{2011}]{VanDerWalt2011}
{Van Der Walt} S.,  Colbert S.~C.,   Varoquaux G.,  2011, \mn@doi [Comput. Sci. Eng.] {10.1109/MCSE.2011.37}, 13, 22

\bibitem[\protect\citeauthoryear{Virtanen et~al.,}{Virtanen et~al.}{2020}]{Virtanen2020}
Virtanen P.,  et~al., 2020, \mn@doi [Nat. Methods] {10.1038/s41592-019-0686-2}, 17, 261

\bibitem[\protect\citeauthoryear{Wakker \& van Woerden}{Wakker \& van Woerden}{1997}]{Wakker1997}
Wakker B.~P.,  van Woerden H.,  1997, \mn@doi [Annu. Rev. Astron. Astrophys.] {https://ui.adsabs.harvard.edu/link_gateway/1997ARA&A..35..217W/doi:10.1146/annurev.astro.35.1.217}, 35, 217

\bibitem[\protect\citeauthoryear{Wakker, Oosterloo  \& Putman}{Wakker et~al.}{2002}]{Wakker2002}
Wakker B.~P.,  Oosterloo T.~A.,   Putman M.~E.,  2002, \mn@doi [Astron. J.] {10.1086/339478}, 123, 1953

\bibitem[\protect\citeauthoryear{Wakker, York, Wilhelm, Barentine, Richter, Beers, Ivezi{\'{c}}  \& Howk}{Wakker et~al.}{2008}]{Wakker2008}
Wakker B.~P.,  York D.~G.,  Wilhelm R.,  Barentine J.~C.,  Richter P.,  Beers T.~C.,  Ivezi{\'{c}} {\v{Z}}.,   Howk J.~C.,  2008, \mn@doi [Astrophys. J.] {10.1086/523845}, 672, 298

\bibitem[\protect\citeauthoryear{Xu}{Xu}{2020}]{Xu2020}
Xu S.,  2020, \mn@doi [Mon. Not. R. Astron. Soc.] {10.1093/mnras/stz3092}, 492, 1044

\bibitem[\protect\citeauthoryear{Zhang, Thompson, Quataert  \& Murray}{Zhang et~al.}{2017}]{Zhang2017a}
Zhang D.,  Thompson T.~A.,  Quataert E.,   Murray N.,  2017, \mn@doi [Mon. Not. R. Astron. Soc.] {10.1093/mnras/stx822}, 468, 4801

\bibitem[\protect\citeauthoryear{von Hoerner}{von Hoerner}{1951}]{vonHoerner1951}
von Hoerner S.,  1951, \mn@doi [Zeitschrift f{\"{u}}r Astrophys.] {https://ui.adsabs.harvard.edu/abs/1951ZA.....30...17V/abstract}, 30, 17

\makeatother
\end{thebibliography}



\appendix

\section{Impact of Simulation Resolution}\label{resolution}

Because simulation resolution has the potential to impact results we present here,  we now investigate this using two simulated clouds in the suite presented by \citet{Abruzzo2024}. Where the previously-shown simulations have a constant resolution of $R_{\rm cl}/\Delta x=16$, the simulations in this Appendix now vary from $R_{\rm cl}/\Delta x=4$ to $R_{\rm cl}/\Delta x=64$. The properties of the clouds are presented in Table~\ref{table:resolution_sim_parameters}, where we note that each cloud (Cloud 100 and Cloud 1000) is referred to based on density contrast values $\chi$. Because these simulations were not introduced by this paper with the intent of attempting to replicate the Smith Cloud, we do not match the GALFA-H\MakeUppercase{\romannumeral1} beam function, add noise, or use noise reduction methods outlined in Section~\ref{simulationinfo}.

\begin{table}
\caption{Initial conditions of each simulation for a brief resolution study, including the density contrast $\chi = \rho_{\rm cl}/\rho_{\rm w}$, radius of the cloud ($R_{\rm cl}$), temperature of the wind ($T_{\rm w}$), and all resolutions included for that particular cloud in terms of $R_{\rm cl}/\Delta x$. Both simulations have thermal pressure $p/k_B=10^3$ K $\rm cm^{-3}$, Mach number $\mathcal{M}=1.5$, and metallicity $Z_{\rm cl}=Z_{\odot}$. Clouds are named according to their density contrast in the first column (e.g., Cloud 100 has $\chi=100$).}
\label{table:resolution_sim_parameters}
    \begin{tabular}{lcccccc} 
\hline
$\chi$ & $R_{\rm cl} \; {\rm [pc]}$ & $T_{\rm w} \; {\rm [K]}$ & $R_{\rm cl}/\Delta x$ \\ 
\hline
100 & 56.38 & $3.28\times10^5$ & 4, 8, 16, 32 \\

1000 & 864.70 & $3.28\times10^6$ & 4, 8, 16, 32, 64 \\
\hline
\hline \\
    \end{tabular}
\end{table}

\begin{figure*}
    \centering
    \includegraphics[width=\textwidth]{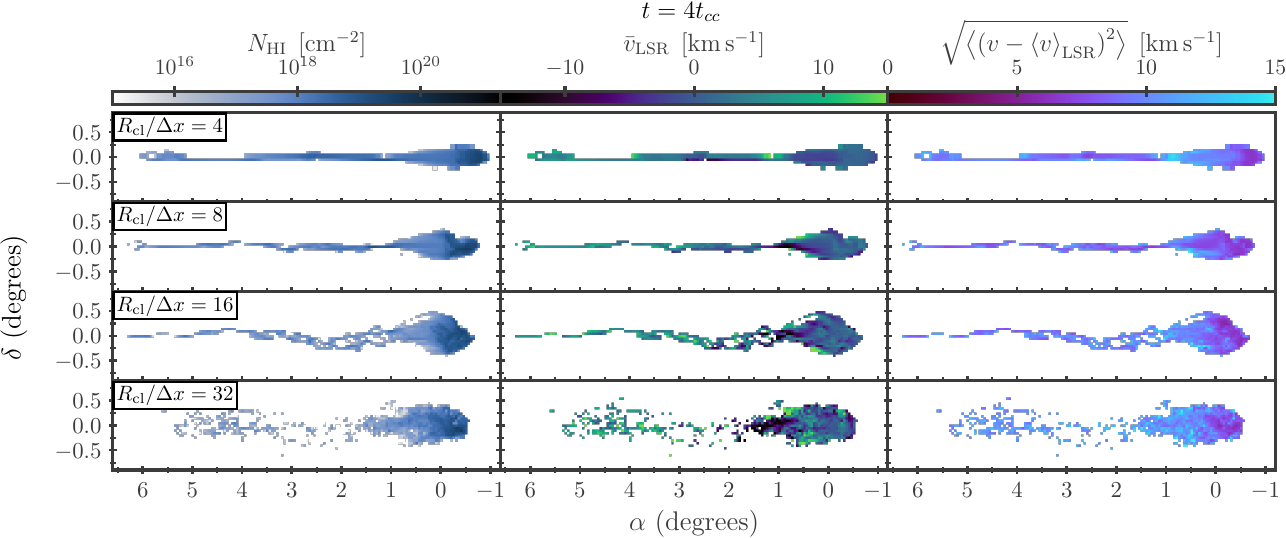}
    \caption{Moment maps of Cloud 100, with rows organized by resolution and columns as each moment, similar to Figures~\ref{fig:sims_tcc03_2sigma} and \ref{fig:sims_tcc12_2sigma}, at $4t_{\rm cc}$. Small-scale structures, such as individual clumps in the cloud-wind interaction, are most identifiable at $R_{\rm cl}/\Delta x=32$. At lower resolutions, the cloud forms a nearly linear tail that is longer than at the highest resolution (bottom row). While subtle, the decreasing column densities in the tail of the cloud, as well as the extent of the velocity variations, are not as evident at lower resolutions.}
\label{fig:X100_resolution}
\end{figure*}

Figure~\ref{fig:X100_resolution} shows the morphology for Cloud 100 over each moment (columns) with increasing resolution (rows) at $t/t_{\rm cc}=4$. It is evident that the small, clump-like structures, often referred to as cloudlets, are most easily seen at the highest-resolution (bottom row). At lower resolutions, unresolved cloudlets appear to form a more linear tail, and deviations in column densities (first column) and $\overline{v}_{\rm LSR}$ (second column) are less obvious. 

\begin{figure*}
    \centering
    \includegraphics[width=0.9\textwidth]{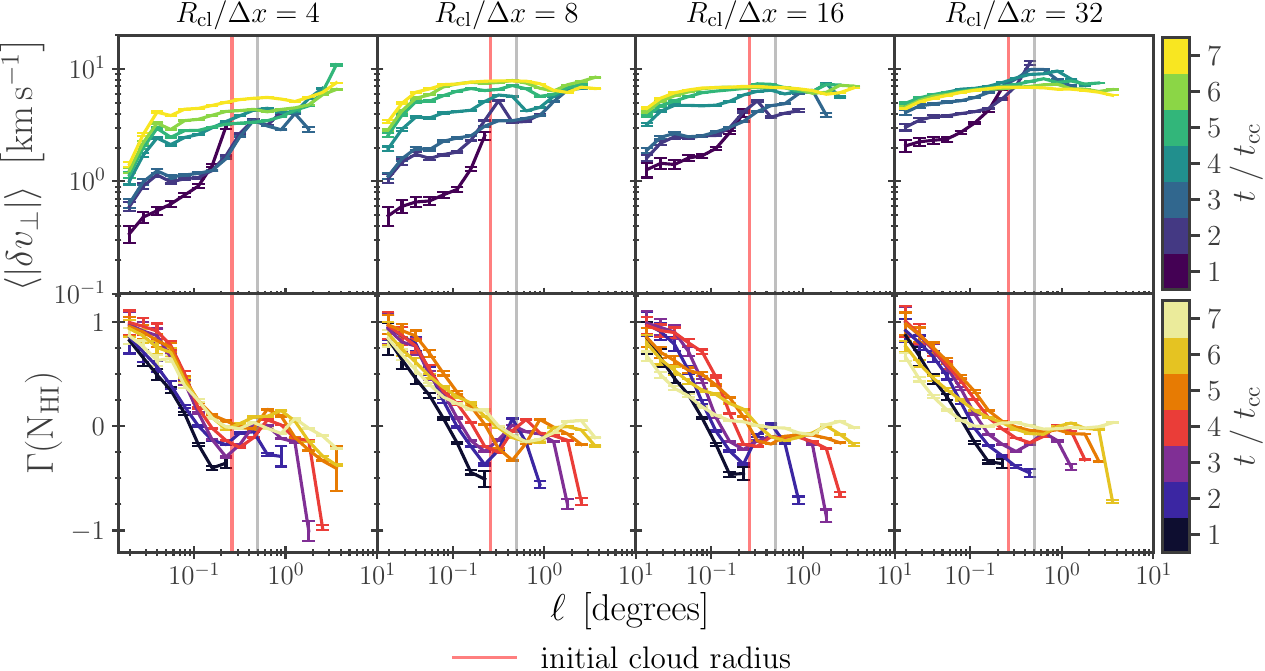}
    \caption{Projected first-order velocity structure function (top row) and normalized autocovariance function of column density (bottom row), colored by cloud-crushing times up to $t/t_{\rm cc}=7$, for Cloud 100. As Figures~\ref{fig:vsf} and \ref{fig:2pc}, the vertical grey line serves as a point of reference for $\ell=0.5^{\circ}$, and the vertical red line is the cloud's initial radius. Resolution increases with each column, with the left column showing the lowest resolution ($R_{\rm cl}/\Delta x=4$) and the right column having the highest resolution ($R_{\rm cl}/\Delta x=32$). Increasing resolution appears to shift the velocity structure function to larger values of $\langle | \delta v_\perp | \rangle$, and results in less variation at later cloud-crushing times. $\rm{\Gamma}(N_{HI})$ tends to flatten slightly with increasing resolution, particularly at later times.}
    \label{fig:VSF_ACF_X100_resolution}
\end{figure*}

The projected velocity structure functions and normalized autocovariance functions of column density for Cloud 100 are both found in Figure~\ref{fig:VSF_ACF_X100_resolution}, colored by $t/t_{\rm cc}$ as Figures~\ref{fig:vsf} and \ref{fig:2pc}. For the VSF, as resolution increases, the function appears to be vertically compressed to higher values and shows less variation. Increased resolution also shows larger values of $\langle |\delta v_{\perp}| \rangle$ at low $\ell$, indicating that turbulence within the cloud is poorly resolved at $R_{\rm cl}/\Delta x=4-8$, and larger resolutions are required to properly these scales. There is less variation in VSF values at high $\ell$, pointing to the ability of the simulation to resolve the bulk velocity of the system.

Deviations in $\rm{\Gamma}(N_{HI})$ are more subtle, but the function appears to be smoother with increasing resolution, most visible at $7t_{\rm cc}$, which is almost completely flat for $\ell>0.1$ at $R_{\rm cl}/\Delta x=32$.

\begin{figure*}
    \centering
    \includegraphics[width=0.9\textwidth]{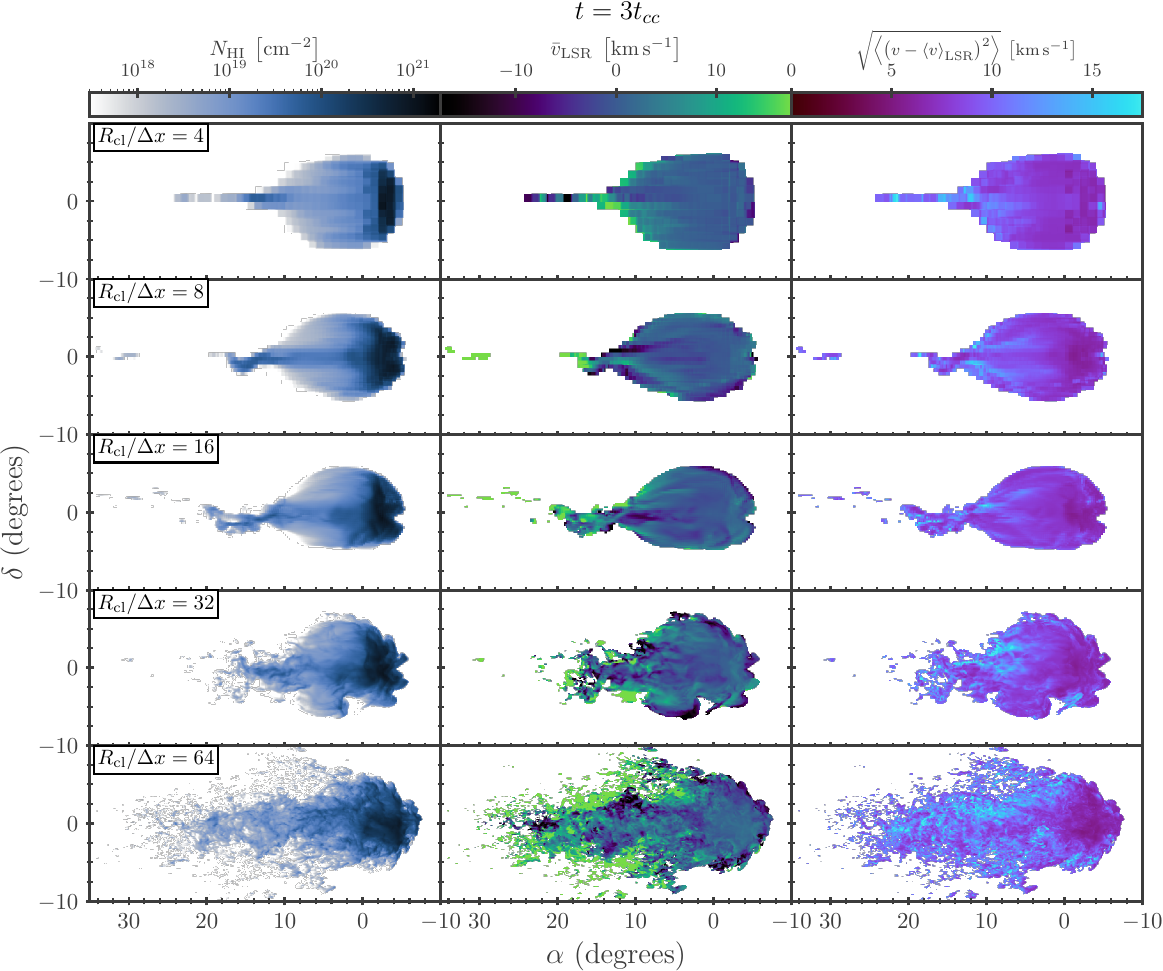}
    \caption{Same as Figure~\ref{fig:X100_resolution}, but for Cloud 1000 at $3t_{\rm cc}$. We note that, from Cloud 100, the minimum column density (first column) colorbar has been increased by about 100 $\rm cm^{-2}$, and mean velocity (second column) has been increased by about $\pm$2.5 km/s. As with Figure~\ref{fig:X100_resolution}, small-scale structures within the cloud become more visible with higher resolution.}
    \label{fig:X1000_resolution}
\end{figure*}

Similarly, Figure~\ref{fig:X1000_resolution} displays the spatial moment distributions of the significantly larger Cloud 1000 at three cloud-crushing times, now with $R_{\rm cl}/\Delta x=64$. At the lowest resolution of $R_{\rm cl}/\Delta x=4$ (top row), the cloud appears to maintain much of its original spherical shape and uniform velocity, with a linear tail (as with Cloud 100 in Figure~\ref{fig:X100_resolution}) being visible. Perturbations become more obvious with increasing resolution, where structure begin to form at the head and tail of the cloud. This is best seen at the highest resolutions (bottom two rows) where the cloud structure becomes much more complex. Again, the lowest resolutions do not capture the same deviations in each property that are visible at the highest resolution.

\begin{figure*}
    \centering
    \includegraphics[width=0.9\textwidth]{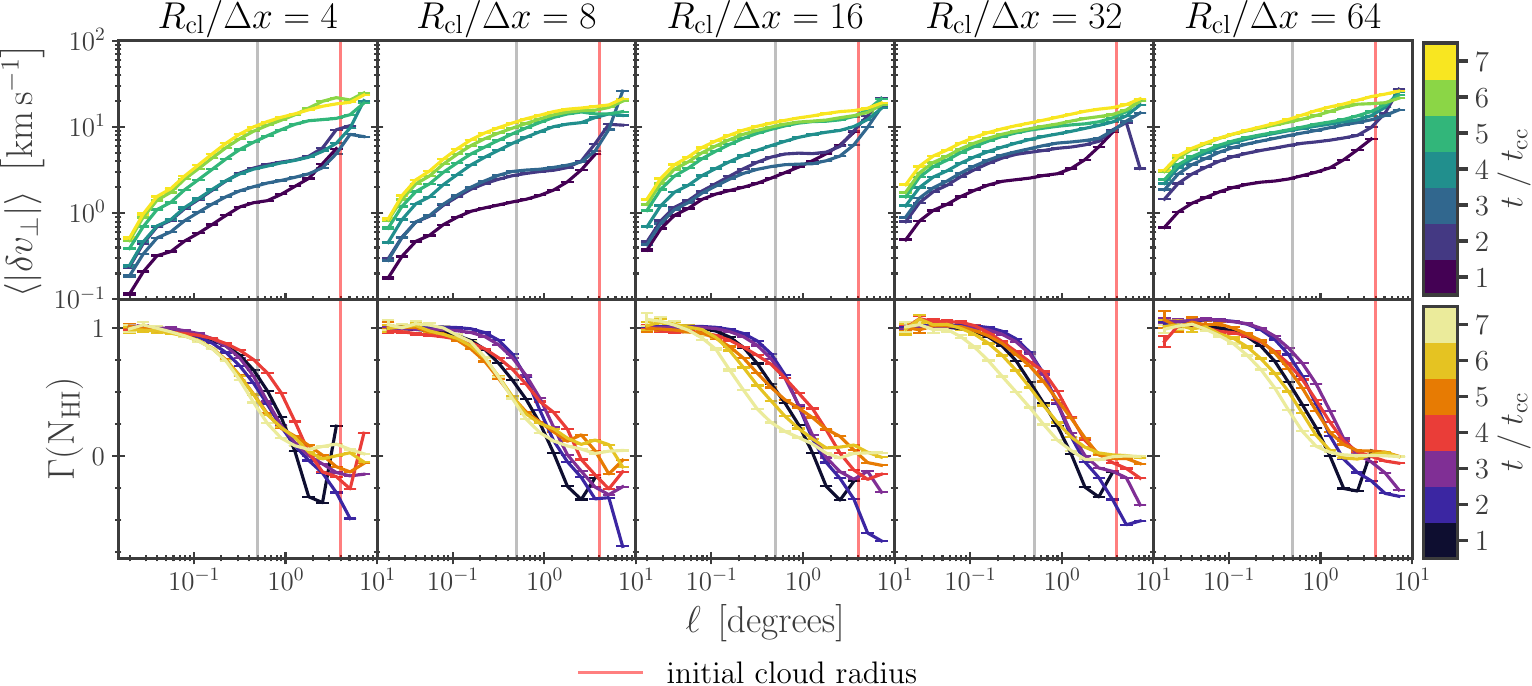}
    \caption{Same as Figure~\ref{fig:VSF_ACF_X100_resolution} for our second resolution study simulation, Cloud 1000, where the highest resolution is now $R_{\rm cl}/\Delta x=64$. Increasing resolution appears to cause the VSF the converge at $t/t_{\rm cc}>1$, increasing to larger values at smaller separations, resulting in the function shifting upwards. In contrast, $\rm{\Gamma}(N_{\rm HI})$ has little difference over time with $R_{\rm cl}/\Delta x=4$.}
    \label{fig:VSF_ACF_X1000_resolution}
\end{figure*}

Finally, Figure~\ref{fig:VSF_ACF_X1000_resolution} is the projected VSF and normalized ACF of neutral hydrogen column density for Cloud 1000. The variations in the VSF with resolution are the same as found in Cloud 100: the VSF appears to be vertically compressed with increasing resolution as values at low $\ell$ are significantly affected . We note that this variation in the projected first-order VSF is likely related to when each respective simulation is able to resolve significant variations in $\overline{v}_{\rm LSR}$, as the VSF is calculated from this parameter; increased resolution means that the simulation is able to better resolve small-scale turbulence.

The normalized autocovariance function of column density, $\rm{\Gamma(N_{\rm HI})}$, appears in the bottom row of Figure~\ref{fig:VSF_ACF_X1000_resolution}.  Again like Cloud 100, Cloud 1000 shows subtle dependence on resolution that is less quantifiable than the VSF. 

While we caution that direct comparisons between the Smith Cloud and Clouds 100 and 1000 cannot be drawn, as they were not simulated with the intent of doing so and are not treated the exact same (with respect to convolution, noise, and noise reduction methods), they may provide further support to our conclusion that the shape of $\rm \Gamma(N_{\rm HI})$ is dependent on \textit{both} (1) whether the cloud survives or not, and (2) the cloud's initial radius. Cloud 100, with an extremely small radius of $R_{\rm cl}\sim56$ pc, has a function that immediately decreases from unity, showing an opposite trend to the Smith Cloud GALFA-HI observations, despite the fact that the cloud is not destroyed by $7t_{\rm cc}$. Cloud 1000, on the other hand, has the largest radius, $R_{\rm cl}\sim 865$ pc, nearly twice that of simulations A and B, and the values of $\rm \Gamma(N_{HI})$ are almost identical to the distribution of the Smith Cloud.

This aligns with predictions by \citet{Cooper2009}, who suggest that radiative clouds are fragmented into smaller cloudlets that have the potential to form a filamentary structure if the cloud survives. Specifically, they predict that increased resolution results in more cloudlets (as visible in Figure~\ref{fig:X100_resolution}) as the Kelvin-Helmholtz instability is further resolved. 

Consequently, higher resolutions are needed to distinguish fine structures in neutral hydrogen, especially at earlier times and in the tail of the cloud. Lower resolution results in more linear clouds with less detail in the velocities. As a result, the VSF is also resolution-dependent: because we suggest that the VSF at small separations is descriptive of turbulence within the cloud, these scales must be adequately resolved, and can influence the shape of the function at low $\ell$.


\bsp	
\label{lastpage}
\end{document}